\documentclass[floatfix,twocolumn,showkeys,amsmath,amssymb]{revtex4}
\usepackage{graphicx,marvosym,fleqn,amsmath,amssymb}
  \usepackage{epsfig,latexsym}
  \usepackage{graphicx}		
  \usepackage{dcolumn}		
  \usepackage{bm}		
  \usepackage{placeins}		
  \begin{document}
\title[Whitening \& face processing]{Does Face Image Statistics Predict a Preferred Spatial Frequency for Human Face Processing?}
  \author{Matthias S. Keil}%
  \email{mats@cvc.uab.es}
  \affiliation{%
Basic Psychology Department,
Faculty for Psychology,
University of Barcelona (UB),
Passeig de la Vall d'Hebron 171,
E-08035 Barcelona (Spain)}%
\begin{abstract}
Psychophysical experiments suggested a relative importance of a narrow band of
spatial frequencies for recognition of face identity in humans.  There exists,
however, no conclusive evidence of why it is that such frequencies are preferred.
To address this question, I examined the amplitude spectra of a large number
of face images, and observed that face spectra generally fall off steeper with
spatial frequency compared to ordinary natural images.  When external face
features (like hair) are suppressed, then whitening of the corresponding
mean amplitude spectra revealed higher response amplitudes at those spatial
frequencies which are deemed important for processing face identity.  The
results presented here therefore provide support for that face processing
characteristics match corresponding stimulus properties.\\
{\bf Kewords:} visual cortex; face recognition; image statistics; whitening; amplitude spectra
\end{abstract}

  %
\maketitle
  %

  \def\figlabel#1{\textit{#1}}
  \def\figitem#1{(\figlabel{#1}) }
  \def\halffigwidth{0.4275}
  \def\figref#1{\ref{#1}}

  %
  %
  \def\eq[#1]{equation~\ref{#1}}
  \def\Eq[#1]{Equation~\ref{#1}}
  %
  %
  \def\s{{ }}
  \def\bs{$\!\!$}
  %
  %
  %
  %
  \def\Cite#1{ (\cite{#1})}	
  %
  %
  %
  \def\Caption[#1][#2][#3]{\caption{\label{#1}\small{{\bf #2.} #3}}}
  \def\rnd{\mathrm{rnd}}
  \def\twoD{2-D}			
  \def\ampspec{\mathcal{A}}		
  \def\formula#1#2{{\begin{equation}\label{#1}#2\end{equation}}}
  \def\indexset{\Omega}			
  %
  \def\Figure#1{Figure~\ref{#1}}
  \def\fig#1{figure~\ref{#1}}
  \def\Fig#1{\Figure{#1}}
  \def\suppfig#1{Supp. Fig.~\ref{#1}}
  \def\Suppfig#1{\suppfig{#1}}
  \def\suppfigure#1{Supp. Fig.~\ref{#1}}
  \def\Suppfigure#1{\suppfigure{#1}}
  \def\suppmethods{Supplementary Methods}
  \def\meanslope{\alpha_\mathrm{mean}}	
  \def\maxslope{\alpha_\mathrm{max}}	
  \def\minslope{\alpha_\mathrm{min}}	
  \def\medianslope{\alpha_\mathrm{med}}	
  %
%
%
%
\section{INTRODUCTION\label{intro}}
%
%
%
%
It has been suggested that the processing of sensory information in the brain
has adapted to the specific signal statistics of stimuli\Cite{Barlow89}.
Such stimulus-specific adaptation is tantamount to taking advantage of statistical
regularities in order to encode the highest possible amount of information
about the signal\Cite{Attneave1954,Linsker1988,BaddeleyEtAl98,NadalBrunelParga1998,Wainwright99}
under various constraints.  The constraints include, for example,
minimizing energy expenditure\Cite{LevyBaxter96,LaughlinEtAl98,Lenny03}, 
minimizing wiring costs between processing units\Cite{LaughlinSejnowski03},
or reducing spatial and temporal redundancies in the input signal%
\Cite{Attneave1954,Barlow61,SrinivasanLaughlinDubs82,Atick92a,HosoyaEtAl05}.\\
In the case of visual stimuli, natural images reveal a statistical regularity
that corresponds to an approximately linear decrease of their amplitude spectra
as a function of spatial frequency when scaling both coordinate axis
logarithmically\Cite{Field87,BurtonMoorehead87}.  This property is equivalent
to strong pairwise correlations between pairs of luminance values\Cite{Wiener1964}.
It has been proposed that visual neurons utilize this statistical property in a way that
cells tuned to different spatial frequencies have equal sensitivities\Cite{Field87}.
Thus, neurons tuned to high spatial frequencies should increase their response gain
such that they achieve the same response levels as low frequency neurons.  This is
the \emph{response equalization hypothesis} (which should be distinguished from the
\emph{decorrelation hypothesis})\Cite{SrinivasanLaughlinDubs82,Atick92a,GrahamChandlerField2006}.
Response equalization (``whitening'') may enhance the information throughput from one
neuronal stage to another by adjusting the output of one stage such that it matches the
limited dynamic range of the successive stage\Cite{GrahamChandlerField2006}.\\
The present article unveils a link between statistical properties of face
images and psychophysical data for the processing of face identity.  The
processing of face identity was found to preferably depend on a narrow spatial
frequency band (about $2$ octaves) from $8$ to $16$ cycles per face
\Cite{TiegerGanz1979,FiorentiniMaffeiSandini1983,HayesMorroneBurr1986,PeliEtAl1994,%
CostenParkerCraw1994,CostenParkerCraw1996,Nasanen1999,OjanpaaNasanen2003}.
However, to the best of my knowledge, no explanation has been offered yet of why
it is that face processing mechanisms in the human brain reveal such a preference.\\
Here I analyzed the amplitude spectra of a large number of face images.
Different types of amplitude spectra were considered - with and without
suppression of external face features (hair, shoulders, etc.).
The spectra were whitened (i.e., ``response''-equalized) according to three
different procedures.  In this way it is demonstrated that the main results
are largely independent of the specific method that was used for whitening:
amplitudes were higher at spatial frequencies around $10$ cycles per
face - but only in those spectra where external face features were suppressed.
Therefore, the effect must have been produced by internal face
features (eyes, mouth, nose).
%
%
%
\section{RESULTS\label{results}}
%
%
%
\begin{figure*}[htbp]
		\figitem{a}\scalebox{0.42}{\includegraphics{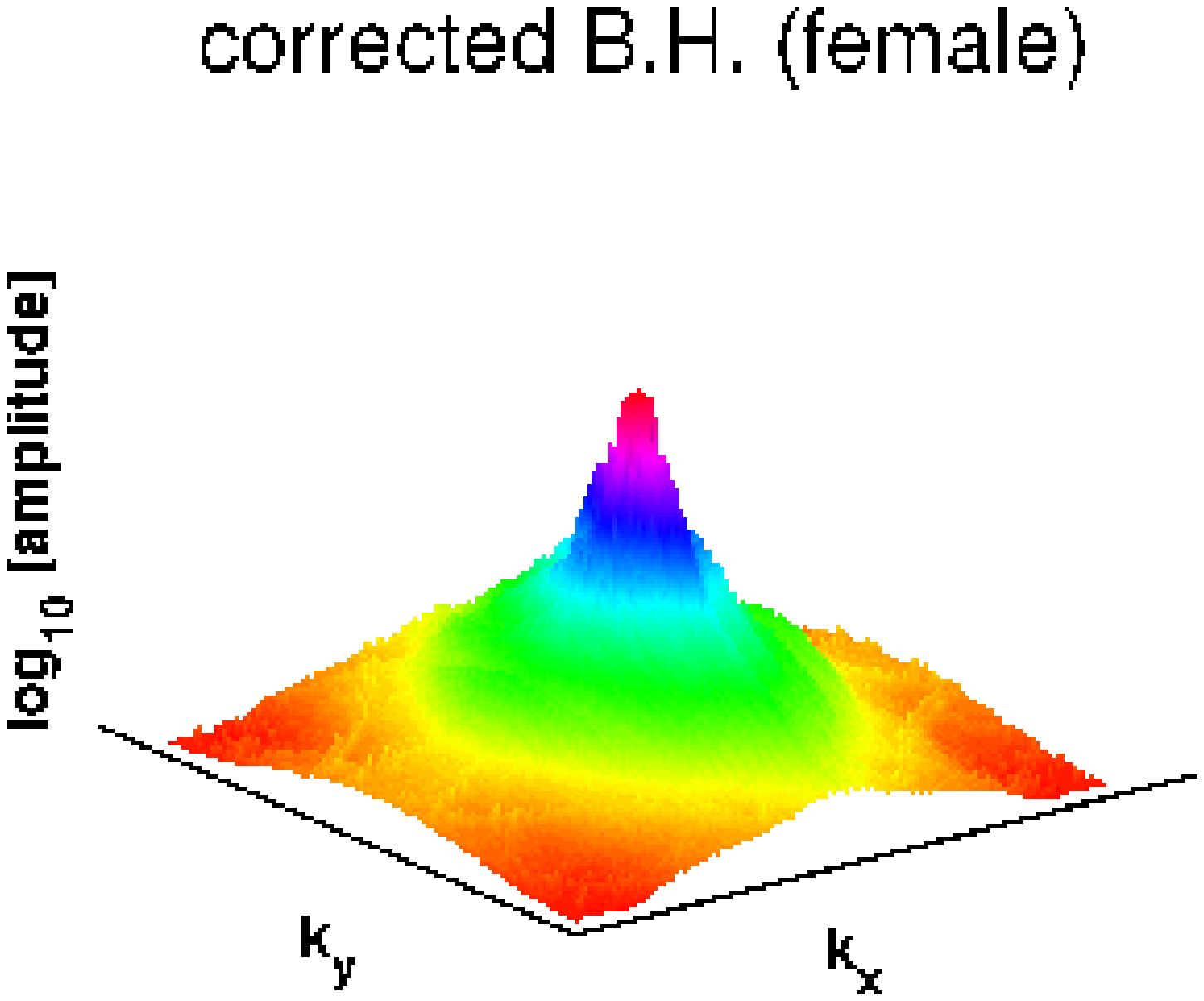}}
		\figitem{b}\scalebox{0.42}{\includegraphics{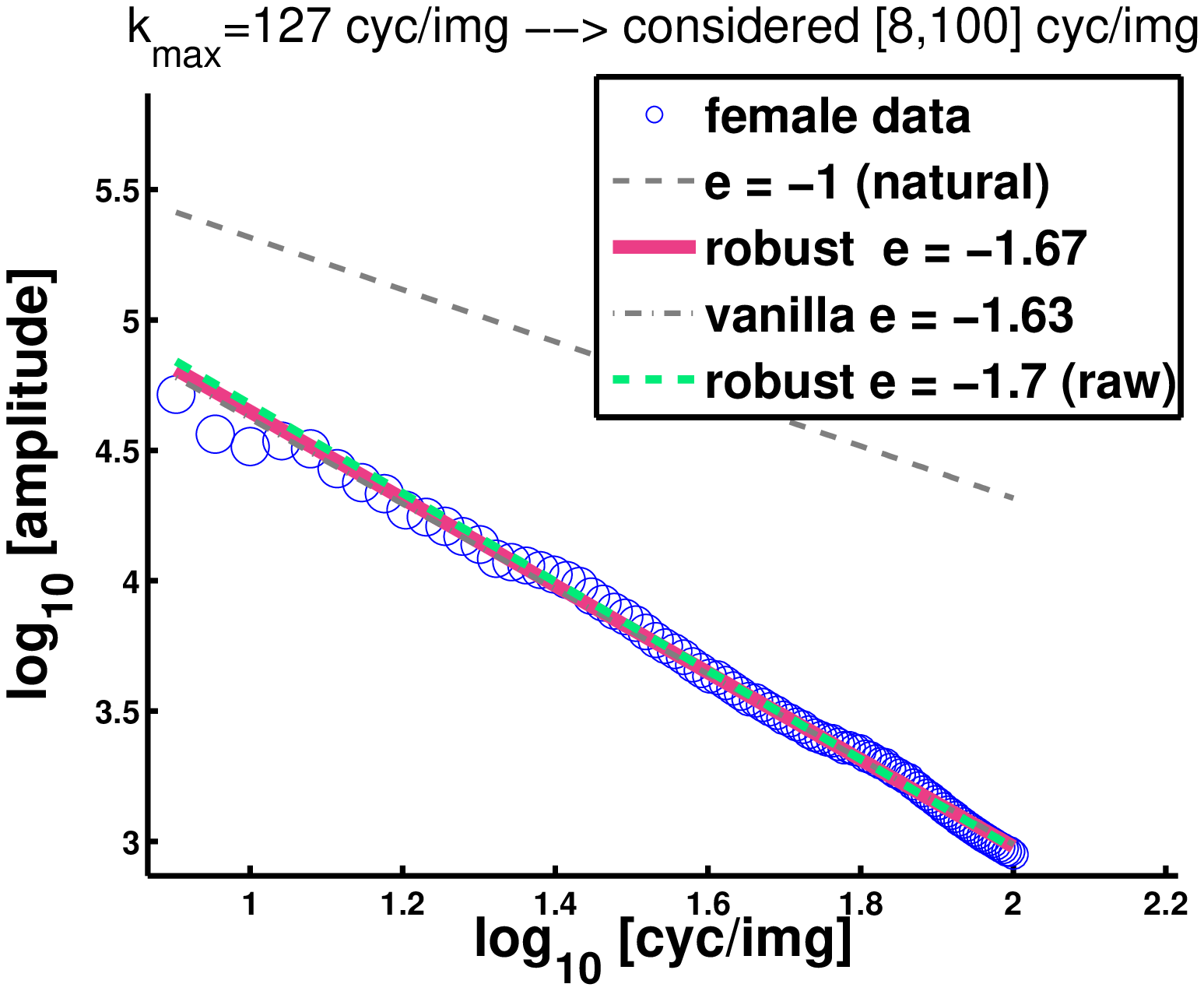}}
	\Caption[corrBH_f][Corrected Blackman-Harris spectrum (females)][{\figitem{a}
	Logarithmized, mean amplitude spectrum of all female face images.  Prior to
	computing individual spectra, a Blackman-Harris (B.H.-) window was applied
	to each face image in order to suppress external face features (\suppfig{MoonFaces}\figlabel{d}).
	The application of the B.H.-window, however, leaves an undesired spectral ``fingerprint''
	in each of the spectra (\suppfig{artifacts}\figlabel{a}), which was attenuated
	before averaging (\suppfig{InwardDiffusionDemo}).\newline
	\figitem{b} The 2-D spectrum shown on the left is transformed into a 1-D isotropic
	spectrum by averaging all amplitudes with different orientations at a fixed frequency
	$k$ ($=$ circle symbols).
	The size of each circle symbol is proportional to the standard deviation (s.d.).
	The maximum s.d. (biggest circle) was $9186.75$ ($39.3\%$), and the minimum s.d.
	(smallest circle) was $252.67$ ($28.08\%$).  In the legend, $e$ denotes slope
	values $\alpha$ (i.e., $\alpha\equiv e$).  For comparison, the typical slope
	of \emph{natural} images ($e=-1$) is also shown as a dashed gray line.
	The label ``\emph{vanilla}'' refers to line fitting with an ordinary linear
	regression (least square fit) algorithm for computing slopes.  Since linear
	regression is sensitive to outliers, slope values were additionally computed
	with an outlier-insensitive ($=$\emph{robust}) algorithm.  Finally, the slope
	for the uncorrected (``raw'') amplitude spectrum is also indicated.}]
\end{figure*}

\def\std#1{\scriptsize{\ensuremath{\pm #1}}}
\begin{table*}
\begin{tabular}{c||c|l|l|l|l}
\textbf{gender}		& \textbf{averaging of:}	& \textbf{raw}		& \textbf{corr.raw}	& \textbf{B.H.}		& \textbf{corr.B.H.}\\
\hline
\hline
female			& \textit{slopes}		& $-1.608$\std{0.0858}	& $-1.604$\std{0.0870}	& $-1.686$\std{0.0698}	& $-1.654$\std{0.0731}\\
\cline{2-6}
			& \textit{spectra}		& $-1.584$		& $-1.579$		& $-1.701$		& $-1.668$\\
\hline
\hline
male			& \textit{slopes}		& $-1.649$\std{0.0738}	& $-1.645$\std{0.0757}	& $-1.673$\std{0.0785}	& $-1.642$\std{0.0895}\\
\cline{2-6}
			& \textit{spectra}		& $-1.644$		& $-1.637$		& $-1.689$		& $-1.658$\\
\hline
\end{tabular}
\caption{\label{SlopeTable}For each gender, the table shows the average slope values for
the four types of amplitude spectra.  Two possibilities for computing these values were considered:
\textit{``slopes''} means that individual slope values were averaged (each gender $n=868$, c.f.
\suppfig{isoslopes_males}), and \textit{``spectra''} refers to the
slope of the average spectrum as illustrated with \Figure{corrBH_f}\figitem{a}.}
\end{table*}
%
  \subsection*{Amplitude spectra}
%
  Amplitude spectra are best conceived in polar coordinates, where the spatial
  frequency $k$ varies proportional to radius.  Thus, spectral amplitudes which
  have the same spatial frequency lie on a circle.  The 2-D spectrum can be collapsed
  into an 1-D isotropic spectrum for each $k$ by averaging all amplitudes on that circle.
  This means that in an isotropic spectrum any orientation dependence of the amplitudes
  is lost.\\
  The amplitude spectra of \emph{natural} images were found to depend on spatial frequency
  as $\propto k^{\alpha}$, with an average (isotropic) spectral slope
  $\alpha\approx-1$\Cite{Field87,BurtonMoorehead87}.\\
  How do the amplitude spectra of face images compare to this finding? 
  To answer, I computed slopes of the amplitude spectra of $868$ female
  and $868$ male face images (size $256\times256$).  In a double-logarithmic representation, these
  spectra also decreased approximately linear as a function of spatial
  frequency (\Figure{corrBH_f}).  Therefore a line with (spectral)
  slope $\alpha$ could be fitted to each spectrum.  Four different
  types of amplitude spectra were considered for each face image (with
  different $\alpha$, see table \ref{SlopeTable} and methods section).\\
  At first the spectra of the original images were computed (``raw').
  The second type of spectrum is defined by attenuating in each spectrum
  the truncation artifacts (``corr. raw'', \suppfig{artifacts}\figlabel{c}
  and \suppfig{InwardDiffusionDemo}).
  These artifacts are a consequence of the cropped shoulder region being displayed
  in each image besides the actual face.  To smoothly
  strip off external face features (like the hair, i.e. anything but the actual face),
  a Blackman-Harris window was applied to each image prior to computing its
  spectrum (``Blackman-Harris'' or ``B.H.'' -- see \suppfig{MoonFaces}\figlabel{d}).
  Because application of the B.H.-window leaves a faint but characteristic
  spectral ``fingerprint'' (\suppfig{artifacts}\figlabel{a}), a further
  spectrum type (``corr. B.H.'') was considered, with the artificial
  ``fingerprint'' being attenuated.\\
  The mean isotropic slope values were computed in two ways.
  First, the spectral slope of each face image was computed, and individual
  slope values were averaged (label \textit{``slopes''} in table \ref{SlopeTable}).
  Second, an average spectrum is computed at first, which is composed of all
  individual spectra (see \Figure{corrBH_f}).  The second slope value corresponds
  then to the slope of the average spectrum  (label \textit{``spectra''} in
  table \ref{SlopeTable}).  Isotropic slope values are situated around $-1.6$,
  with minima and maxima of $-2.014$ \& $-1.180$ (females), respectively, and
  $-1.994$ \& $-1.007$ (males).\\
  Notice that the standard deviations associated with the slopes of arbitrary
  natural images are usually bigger\Cite{TolhurstEtAl92,SchaafHateren96},
  as there is no restriction on displayed content and scale,
  respectively\Cite{TorralbaOliva03}.\\
\begin{figure}[tbph!]
	\scalebox{\halffigwidth}{\includegraphics{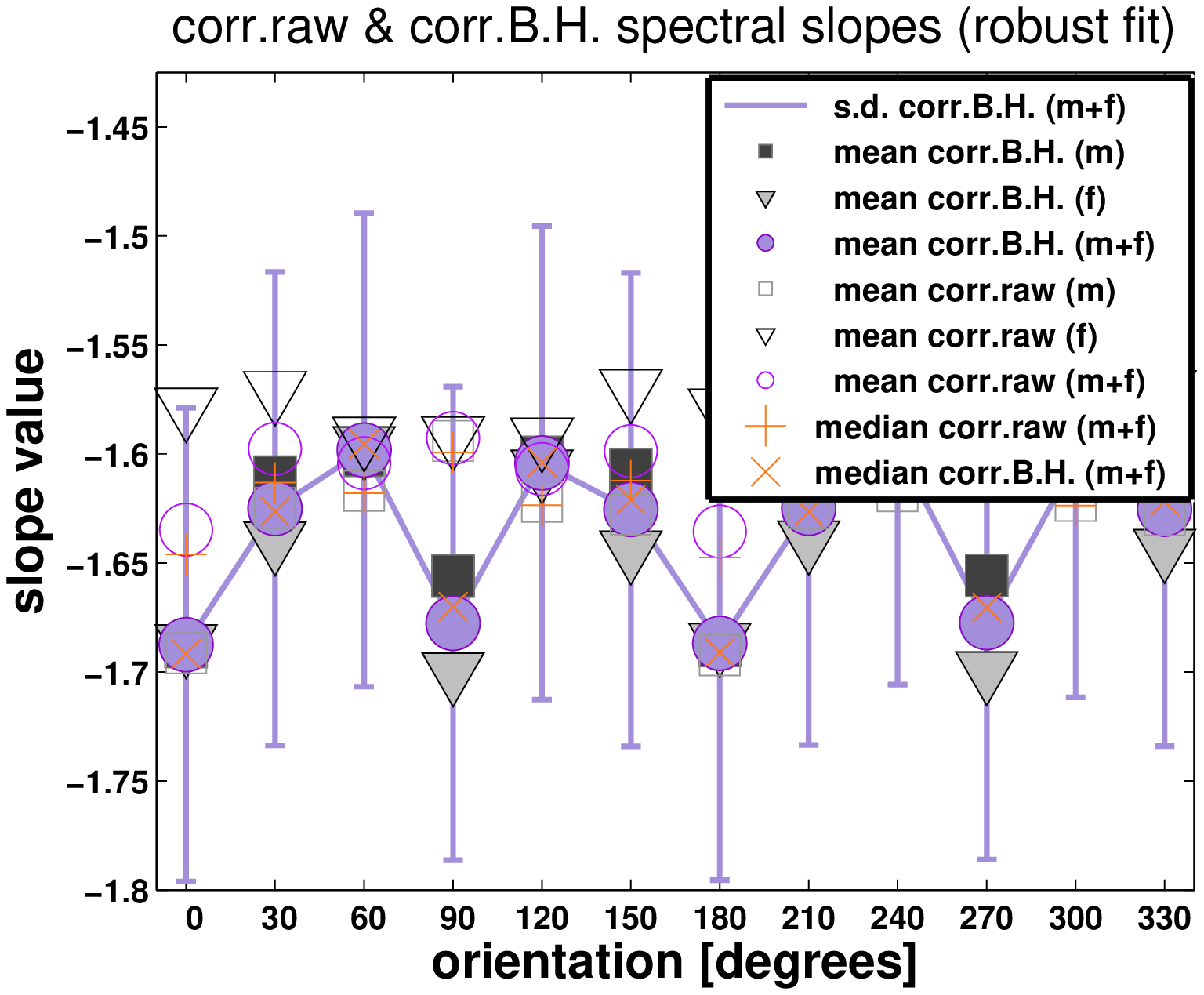}}
	\Caption[AngularSpectralSlopes][Oriented spectral slope][{The
	curves juxtpose oriented spectral slopes from corrected raw
	(``corr.raw'') spectra and corrected Blackman-Harris-windowed
	(``corr.B.H.'') spectra (see legend).  Slopes were computed
	from the respective averaged spectra, with angular increments
	of $30^o$ \Cite{SchaafHateren96}.  Error bars denote
	$\pm1$ standard deviation (estimated using robust statistics).
	Uncorrected spectra show similar dependencies of slopes from
	orientation.  Notice that slope values are defined modulus $180^o$.}]
\end{figure}
  Usually, $\alpha$ varies also as a function of orientation $\Theta$\Cite{SwitkesEtAl78,SchaafHateren96}.
  The orientation dependence is illustrated by means of the averaged corrected spectra (\Figure{AngularSpectralSlopes}).
  Minimum slope values are located at $0^o$ (wave vector pointing to east) and $90^o$
  (north), respectively, whereas maxima tend to be at oblique orientations.  Slope
  values of the B.H. spectra vary more than with the raw spectra.  As external features
  are widely suppressed in the B.H. spectra, minimum slopes are associated with the
  orientations of the internal face features ($0^o,180^o$: nose; $90^o, 270^o$: eyes,
  mouth, and the bottom termination of the nose).\\
  Summarizing so far, the majority of the individual $\alpha$ for face images
  is more negative than the theoretically predicted lower bound of $-1.5$ for natural
  images\Cite{BalboaEtAl01} (table \ref{SlopeTable}; \suppfig{isoslopes_males}).
  Similar observations also hold for spectral slopes
  of the mean amplitude spectra (\Figure{corrBH_f}).
  This should not come as a surprise since the structure of face images is different
  from natural images: face images are not composed of self-occluding, constant
  intensity surface patches\Cite{Ruderman97,BalboaEtAl01}, and lack the self-similar
  distribution of spectral energy as it was reported for natural images\Cite{Field87}.
\begin{figure}[tbph!]
	\scalebox{\halffigwidth}{\includegraphics{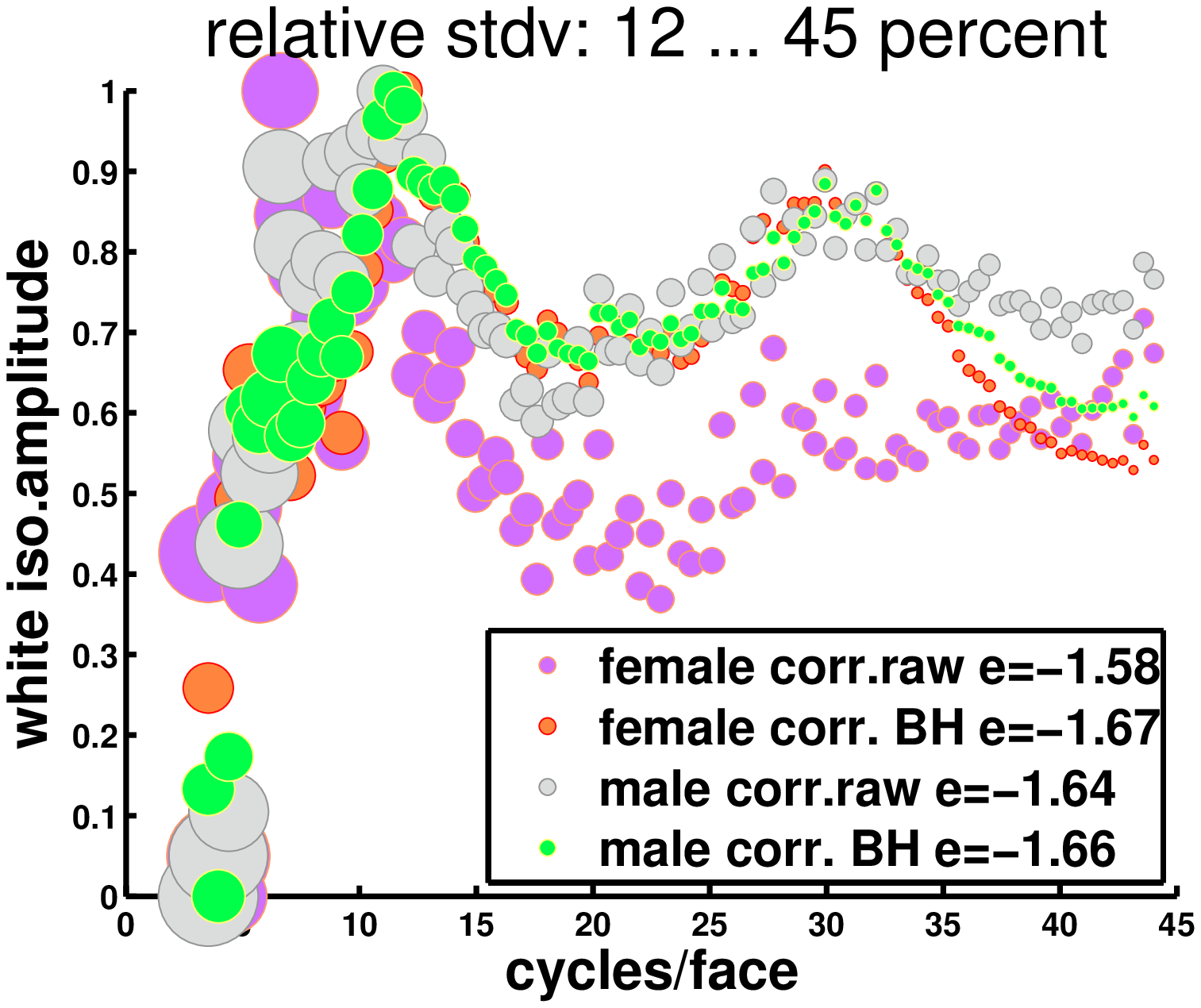}}
	\Caption[IsotropicWhitening][1-D whitening][{Whitening of the
	corrected mean, isotropic 1-D spectra reveals a global amplitude maximum
	at $\approx 10$ cycles per face with all four spectra (see legend).  Symbol
	size is proportional to standard deviation (relative values are indicated
	in the figure).  The slopes which were used for whitening are indicated in
	the legend (c.f. table \ref{SlopeTable}).}] 
\end{figure}
\begin{figure*}[bpth!]
	  \figitem{a}\scalebox{0.425}{\includegraphics{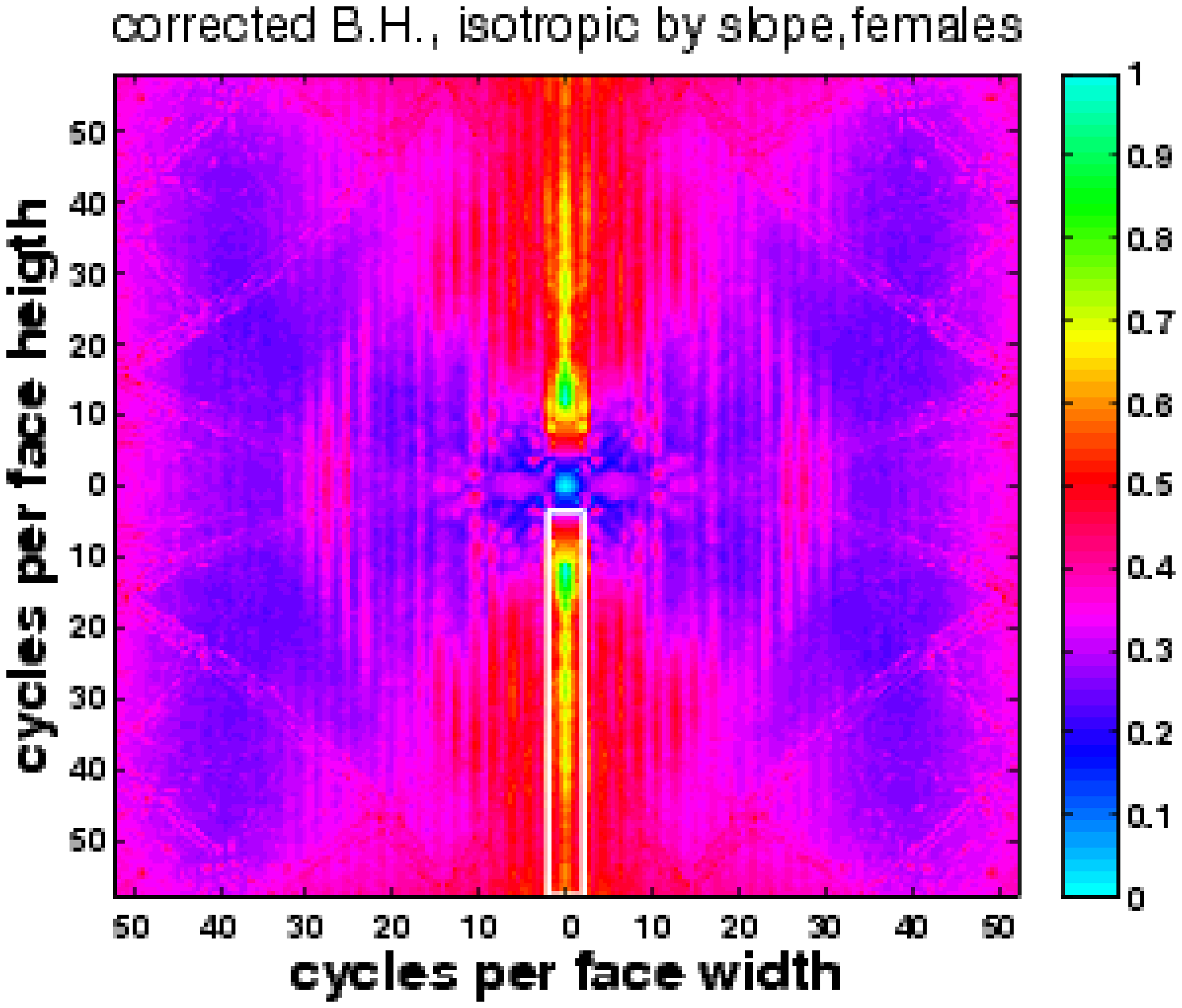}}
	  \figitem{b}\scalebox{0.425}{\includegraphics{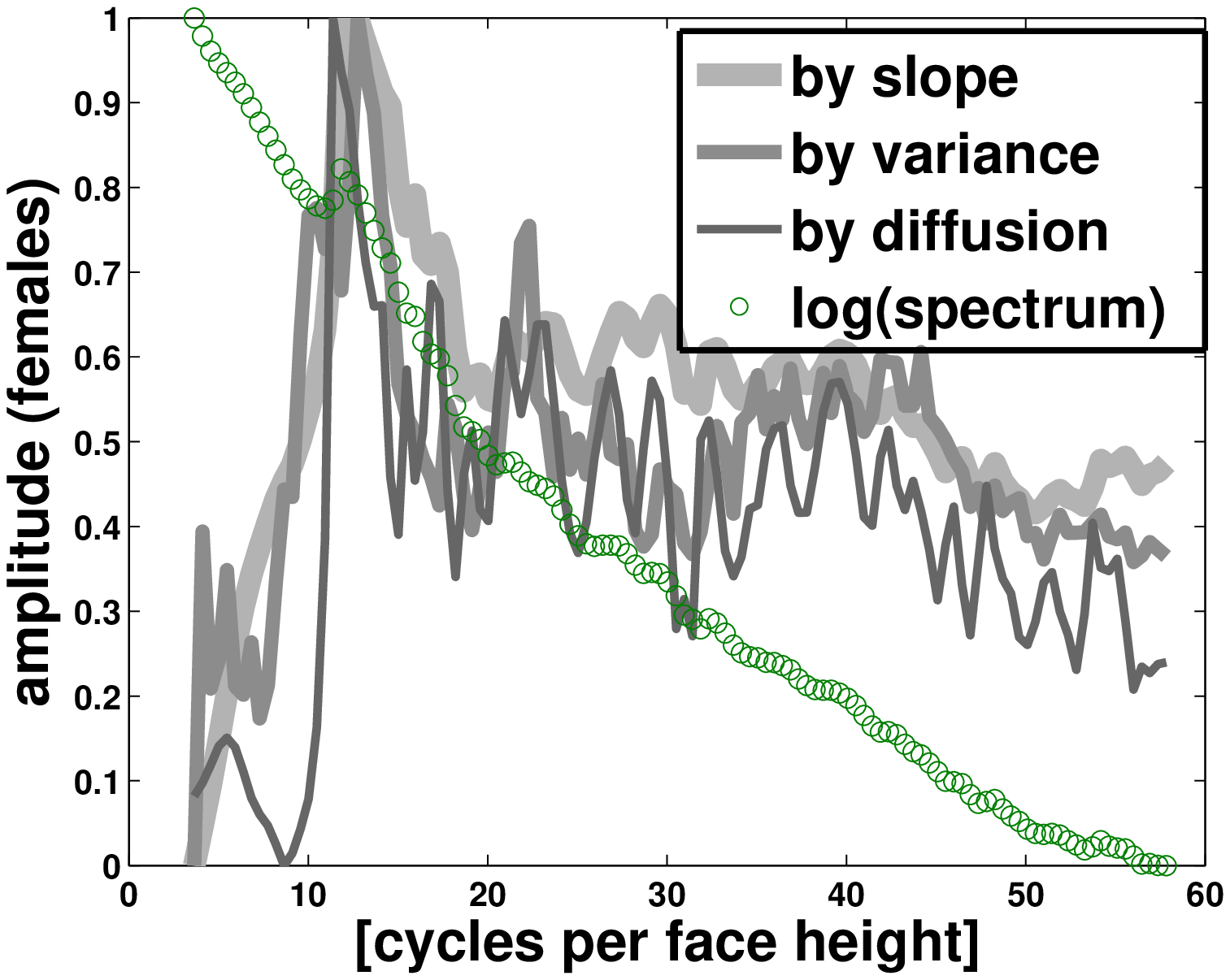}}
	  \Caption[WhiteFemales2D][2-D whitening][{\figitem{a}Slope-whitening of the
	  mean corrected B.H.-spectra unveiled clear maxima at horizontal feature
	  orientations (marked by a white box).  Here the female data are shown
	  (male data: \suppfig{WhiteMales2D}).
	  \figitem{b} The curves show the amplitudes at the location demarcated by
	  the white box in the spectrum: \textit{``log(spectrum)''} are the logarithmized
	  amplitudes without whitening; amplitudes whitened \textit{``by slope''},
	  \textit{``by variance''}, and \textit{``by diffusion''} (see methods sections
	  for further details on the three whitening procedures).  The important result here
	  is that whitened amplitudes reveal a distinct maxima irrespective of the specific
	  whitening method at $\approx 10-15$ cycles per face height.
	  The variance-whitened spectra are shown in \suppfig{VarWhitening}.}]
\end{figure*}
%
%
  \subsection*{Whitening the Amplitude spectra}
%
  Here I ask whether by amplitude equalization of amplitude spectra
  (``whitening'') one could explain psychophysical data on face perception.
  The results which are presented below were obtained with the mean spectra.\\
  Consider first the isotropic (1-D) spectra.  Because the spectra fall,
  as a function of spatial frequency $k$, as  $\propto k^{-|{\alpha}|}$,
  we can multiply amplitudes by $k^{|{\alpha}|}$ to obtain a ``flat''
  spectrum (in the sense that its Shannon entropy is maximal).  The slopes
  which were used to this end are the \textit{``spectra''} ones from table
  \ref{SlopeTable}.  Whitened 1-D spectra are shown in \Figure{IsotropicWhitening}.
  They are not completely flat, but instead have a global maximum at around $10$
  cycles per face, and a second but local maximum at around $30$ cycles per face.\\
  Consider now the 2-D spectra, where whitening was carried out according
  to three different procedures: whitening by \emph{slopes} (analogous to the 1-D case),
  by \emph{variance}, and by \emph{diffusion} (see methods section).  Results are shown
  in \Figure{WhiteFemales2D} for females, and in \suppfig{WhiteMales2D} for males.
  For both genders, the whitened B.H.-spectra reveal amplitude maxima only within
  a narrow band of low spatial frequencies.  Furthermore, frequency maxima appear
  only at a specific orientation in the spectra which corresponds to horizontally
  oriented face features (``horizontal amplitudes'', i.e. eyes and mouth).
  These results are obtained independently from the specific whitening procedure
  which was used (\emph{slope}-whitening: \Figure{WhiteFemales2D}\figlabel{a} \&
  \Figure{WhiteMales2D}\figlabel{a}; \emph{variance}-whitening: \suppfig{VarWhitening};
  \emph{diffusion}-whitening: not shown).\\
  Plotting of only these ``horizontal amplitudes'' (indicated by a white box
  in \Figure{WhiteFemales2D}\figlabel{a}) for all three whitening procedures
  allows to identify the spatial frequencies of the maxima with higher precision.
  The curves now show clearly that the maxima occur in the range from  $10$ to $15$
  cycles per face height.
  Nevertheless, maxima are only revealed by whitening of the B.H.-windowed
  spectra, but not by whitening of any raw spectra.  This means that amplitude
  enhancement due to internal face features is annihilated by the presence
  of external face features (such as hair or shoulder).
%
%
%
%
%
 \section{DISCUSSION\label{discussion}}
%
%
%
 Here, I studied amplitude spectra of face images in the context of response
 equalization (whitening).  Were external face features (hair, shoulder)
 suppressed by windowing the face images with a Blackman-Harris window,
 then amplitude maxima are observed in the whitened spectra at low spatial
 frequencies.  For the isotropic 1-D spectra, maxima are situated around
 $10$ cycles per face, and for the 2-D spectra at around $10-15$ cycles
 per face height.  In the 2-D case, three different whitening methods
 yielded consistent results.\\
 Several psychophysical studies suggest that recognition of face identity 
 works \emph{best} in a narrow band (bandwidth about $2$ octaves)
 of spatial frequencies from $\approx 8$ to $\approx 16$ cycles per
 face\Cite{Ginsburg1978,TiegerGanz1979,FiorentiniMaffeiSandini1983,%
 HayesMorroneBurr1986,PeliEtAl1994,CostenParkerCraw1994,Nasanen1999}.
 Notice that this does not mean that face recognition exclusively depends
 on this frequency band, as faces can still be recognized when corresponding
 frequency information is suppressed\Cite{Nasanen1999,OjanpaaNasanen2003}.\\
 Because the amplitude maxima appear in the whitened spectra exclusively at
 horizontal feature orientations, my results suggest that the psychophysical
 frequency preference might have been caused by an adaptation of corresponding
 neuronal mechanisms to eyes and mouth.\\
Interestingly, in the earlier cited psychophysical studies the spatial frequencies
are often measured in ``cycles per face width'' (i.e., along vertically oriented
face features), whereas the results presented here were rather brought about by
horizontally oriented face features.  The factors to convert spatial frequencies
from ``cycles per image'' to ``cycles per face'' (see methods) are statistically
different for width and height (as suggested by a one-way ANOVA and a Kruskal-Wallis test).
However, they are not so different in absolute terms.  The aforementioned
frequency interval of $10-15$ \emph{cycles per face height} transforms into
$\approx 9-13.5$ \emph{cycles per face width} for females and $\approx 9-13.6$
\emph{cycles per face width} for males, respectively, what is still in good
agreement with the psychophysical data.\\
 Psychophysical thresholds for face recognition are not significantly
 affected by the structure of the background in which a face is
 embedded\Cite{CollinWangOByrne2006}.  Therefore, although the faces
 used in this study are shown against an uniform background, the validity
 of results should extend to arbitrary backgrounds.  Notice, however,
 that amplitude spectra consider the complete frequency content of an
 image, whereas humans have attentional mechanisms which allows them
 to process only a region of interest, and ignore background effects.
 Windowing the face images with a Blackman-Harris window achieves the
 the same computational purpose: anything but the internal face features
 are suppressed.  A follow-up paper examines in more detail the
 properties of internal face features by means of a model of simple
 and complex cells.\\
 The statistical prediction of a preferred band of spatial frequencies
 may also have implications for artificial face recognition systems.  Future
 experiments should systematically address the question whether the recognition
 performance of artificial systems is optimal at spatial frequencies
 similar to those used by humans.
%
%
%
\section*{Methods}
%
%
%
%
\def\firstmethodsec#1{\vspace*{-0.5cm}{\bf #1.}  }
\def\methodsec#1{\vspace*{0.35cm}\\{\bf #1.}  }
\linespread{1.0}
\footnotesize
\firstmethodsec{Face images}\label{faceimgs}
%
%
We used $868$ female face images, and $868$ male face images from the
\emph{Face Recognition Grand Challenge} database (FRGC,
\texttt{http://www.frvt.org/FRGC} or \texttt{www.bee-biometrics.org}).
Original images ($1704\times 2272$ pixels, 24-bit true color) were adjusted for
horizontal alignment of eyes, before they were down-sampled to $256\times256$
pixels and converted into 8-bit grey-scale.  Subsequently,
the positions of left eye, right eye, and mouth [$(x_\mathit{le},y_\mathit{le})$,
$(x_\mathit{re},y_\mathit{re})$, and $(x_\mathit{mouth},y_\mathit{mouth})$,
respectively] were manually marked by two persons (M.S.K. and E.C.) with
an \emph{ad hoc} programmed graphical interface.
The position of each face center ($\approx$ nose) was approximated as
$x_\mathit{nose}=\rnd((x_\mathit{le}+x_\mathit{re})/4  + x_\mathit{mouth}/2)$ and
$y_\mathit{nose}=\rnd [0.95*\rnd (y_\mathit{le}  + (y_\mathit{mouth}- (y_\mathit{le}+y_\mathit{re})/2)/2 )$,
where $\rnd(\cdot)$ denotes rounding to the nearest integer value.
%
\methodsec{Dimension of spatial frequency}\label{SpatialFrequencyConversion}
%
%
For conversion of spatial frequency units, face dimensions were manually
marked with an \emph{ad hoc} programmed graphical interface.
The factors for multiplying ``cycles per image'' to obtain ``cycles per
face width'' were $0.41\pm 0.013$ (females, $n=868$) and $0.43\pm0.012$
(males, $n=868$).  Corresponding factors for obtaining ``cycles per face 
height'' were $0.46\pm0.021$ (females) and $0.47\pm0.018$ (males).
Conversion factors at oblique orientations were calculated
under the assumption that horizontal and vertical conversion factors
define the two main axis of an ellipse.  Pooling of results over gender
implied also a corresponding averaging of conversion factors, and
the factors for width and height were averaged in the isotropic case.
%
\methodsec{Amplitude spectra}\label{amplitude_spectra}
%
Let the features which are not part of the actual face be denoted
by \emph{external features} (e.g., shoulder region or hair).  On the other
hand, \emph{internal features} refer to the eyes, the mouth, and the nose. 
The presence of external features in our face images influences in their
amplitude spectra, and may cause truncation artifacts.
It is thus desirable to compare results with and without the
presence of external features.  A good suppression of external features
could be achieved by centering a \emph{minimum 4-term Blackman-Harris window}\Cite{Harris78}
at $(x_\mathit{nose},y_\mathit{nose})$ (\suppfig{MoonFaces} \& \figref{fems_testMI}).
Nevertheless, application of the window leaves a characteristic ``fingerprint''
in each spectrum (\suppfig{artifacts}\figlabel{a}).  This artificial ``fingerprint'',
as well as the spurious lines caused by truncation, could be attenuated
with a correction procedure based on a spatially varying diffusion mechanism (outlined below).
Thus, for each face image, originally four types of amplitude spectra were considered: 
the original ``raw'' spectrum, the ``Blackman-Harris''-spectrum,
and their respective corrected versions (i.e., ``corr. raw'' and ``corr. B.H.'').
%
%
\methodsec{Correction of Amplitude Spectra}\label{CorrectingAmplitudeSpectra}
%
%
Let $P\in\{0,1\}^{n \times n}$ be a binary $n\times n$ matrix of the same size
as the 2-D amplitude spectra $\mathcal{A}$.  In $P$, artifacts are represented
by ones, while all other positions are set to zero.  Thus, $P$ is set to the
image shown in \suppfig{artifacts}\figitem{b} for correcting the Blackman-Harris
spectrum, and \suppfig{artifacts}\figitem{c} for the raw spectrum.  The idea of
the correction algorithm consists in simply averaging out the positions with artifacts.
To this end, information from neighboring positions flows into artifact positions.
This process is called \emph{inward diffusion}.
Let $\mathcal{X}(t)$ be a sequence of corrected amplitude spectra parameterized over
time $t$, with the initial condition $\mathcal{X}(0)\equiv\mathcal{A}$.  Inward
diffusion is defined by
$\partial \mathcal{X_\mathrm{ij}}/ \partial t = P_\mathrm{ij}\nabla^2 \mathcal{X_\mathrm{ij}}$,
where $(i,j)$ denotes matrix positions.  The diffusion process was terminated at the moment
when the \emph{correlation difference} $c(t)-c(t+\Delta t)$ was smaller than $0.001$, or when a maximum
of $100$ iterations were done.
%
%
\methodsec{Slopes of amplitude spectra}\label{ComputingAmpSlopes}
%
%
\emph{Isotropic slopes $\alpha$:} Amplitudes associated with a given spatial
frequency lie on a circle.  This is to say that when representing the spectrum
with polar coordinates, then spatial frequencies vary along the radial coordinate,
but stay constant while varying orientation.
An isotropic amplitude spectra is obtained by averaging all amplitudes with
a fixed spatial frequency across orientations (i.e., for each circle, the mean
value of all amplitudes of the circle was computed).  Because the logarithmized
amplitude spectra of face images fall approximately linear as a function of
log-frequency, a line with slope $\alpha$ could be fitted to the isotropic spectra.
Although in principle amplitude data were available from $k=1$ to $k=127$ cycles
per image, only the interval from $k_{min}=8$ to $k_{max}=100$ was used for
line fitting.  I used the function  ``\texttt{robustfit}'' (linear regression
with low sensitivity to outliers) provided with Matlab's statistical toolbox
(\emph{Matlab} version 7.1.0.183 R14 SP3, \emph{Statistical Toolbox} version 5.1,
see \texttt{www.mathworks.com}).\\
\emph{Oriented spectral slopes $\alpha(\Theta)$} (\Figure{AngularSpectralSlopes}):
Each 2-D amplitude spectrum was subdivided into $12$ ``pie slices'' (each with
$\Delta\Theta=30^o$).  For each pie slice with orientation $\Theta$, an (oriented)
isotropic 1-D spectrum was analogously computed as just described (with amplitudes
being averaged across arcs), and subsequently a line with slope $\alpha(\Theta)$
was fitted.
%
%
\methodsec{Slope-Whitening of Amplitude Spectra}\label{SlopeWhitening}
%
%
This algorithm proceeds in straight analogy to whitening of the isotropic spectra.
Let $\alpha$ be the isotropic slope value corresponding to a 2-D amplitude
spectrum $\mathcal{A}(k_x,k_y)$ with spatial frequency coordinates $k_x$,
$k_y \in [1,127]$ cycles per image.  Let $k=\sqrt{k_x^2+k_y^2}$ (radial spatial
frequency).  Then, the corresponding whitened spectrum  $\mathcal{W}$ is defined as 
$\mathcal{W}(k_x,k_y)=\mathcal{A}(k_x,k_y) \cdot k^{|\alpha|}$.  Qualitatively, the
$\mathcal{W}$ were not different from a more advanced procedure that consisted in
subdividing $\mathcal{A}$ into oriented ``pie slices'' and whitening each
with its corresponding oriented slope value $\alpha(\Theta)$.  Therefore, only those
results are presented where $\mathcal{A}$ was whitened with an isotropic slope value
(the term ``isotropic'' in the headline of the spectra in \Figure{WhiteFemales2D}
and \suppfig{WhiteMales2D} indicates just this).
%
%
%
\methodsec{Whitening by Variance}\label{VarianceWhitening}
%
%
Amplitudes in the spectrum $\mathcal{A}(k_x,k_y)$ with equal spatial frequencies lie
on a circle with radius $k=\sqrt{k_x^2+k_y^2}$.  Let $n_k$ be the number of points on
this circle ($n_k$ monotonically increases as a function of $k$).  Let  $\mathcal{A}(k,\Theta)$
be the spectrum in polar coordinates.  Then, we first average, for each $k$, all amplitudes
across orientations according to $\mu(k)=\sum_{\Theta}\mathcal{A}(k,\Theta)/n_k$.  The
variance is subsequently computed as $\sigma^2(k)=\sum_{\Theta}(\mathcal{A}(k,\Theta)-\mu)^2/(n_k-1)$.
Finally, the variance-whitened spectrum is defined as
$\mathcal{V}=\mathcal{A}/(\sigma^2(k)+ \epsilon)$ with a small positive
constant $\epsilon\ll 1$.  Examples of $\mathcal{V}$ are shown in \suppfig{VarWhitening}.
%
%
\methodsec{Whitening by Diffusion}\label{DiffusionWhitening}
%
%
Let $\mathcal{X}(k_x,k_y,t)$ a sequence of amplitude spectra parameterized over
time $t$, with the initial condition $\mathcal{X}(k_x,k_y,0)\equiv\mathcal{A}(k_x,k_y)$.
For $t>0$, the $\mathcal{X}$ are defined according to the diffusion equation
$\partial \mathcal{X}/ \partial t = \nabla^2 \mathcal{X}$.  The whitened spectrum
then is $\mathcal{D}\equiv \mathcal{A}/(1+\mathcal{X}(t_\mathrm{max}))$
at precisely the instant $t_\mathrm{max}$ when the Shannon entropy of
$\mathcal{D}$ is maximal.
%
%
%
\normalsize
\linespread{1.7}
%
 \section*{Acknowledgements}
 This work was partially supported by the \emph{Juan de la Cierva}
 program from the Spanish Government (BKC-IYK-6707).  Further support
 was granted by the MCyT grant SEJ 2006-15095.  M.S.K. wishes to thank
 Esther Calder{\'o}n for her valuable help in acquiring feature positions,
 as well as Hans Sup{\`e}r for helpful comments.
\clearpage
\section*{\Large Supplementary Figures}
\setcounter{page}{1}
%
\begin{figure*}[htbp]
		  \figitem{a}\scalebox{0.5}{\includegraphics{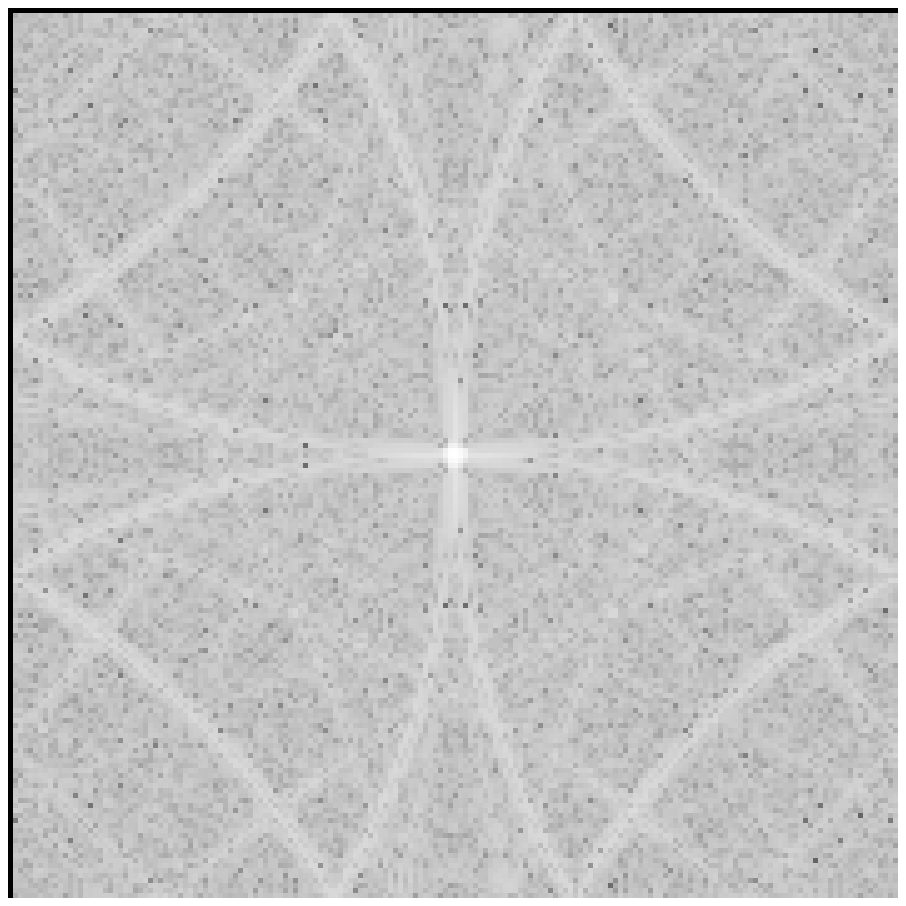}}
		  ~~\figitem{b}\scalebox{0.5}{\includegraphics{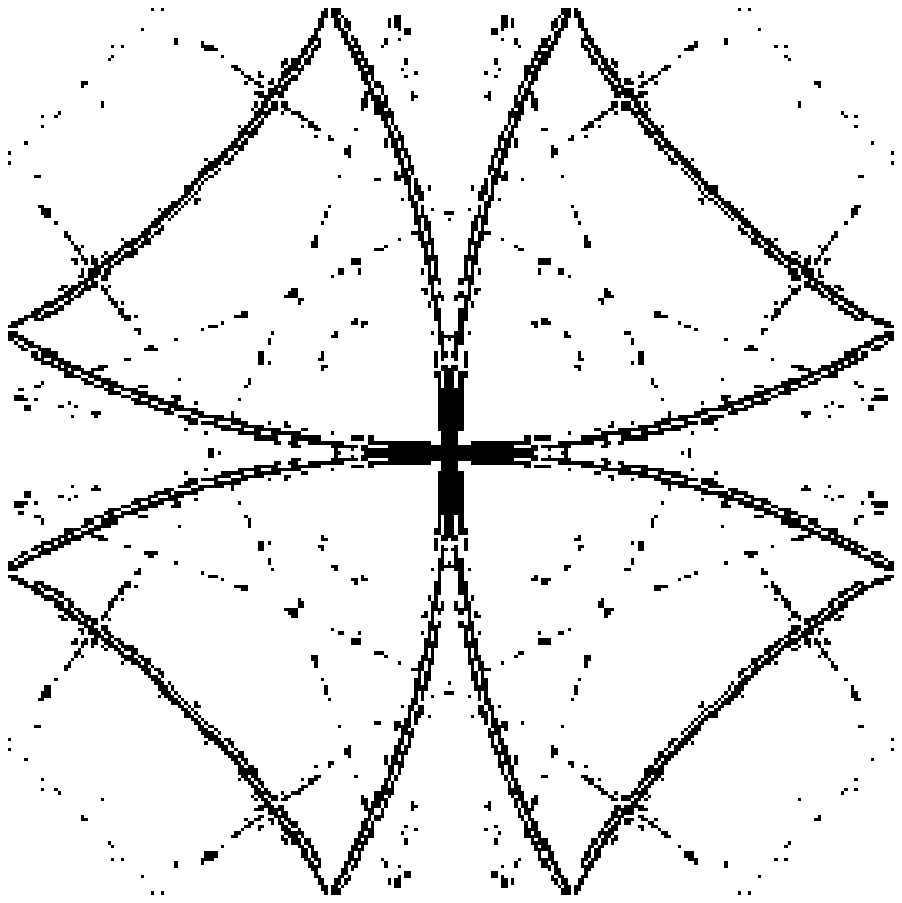}}
		  ~~\figitem{c}\scalebox{0.5}{\includegraphics{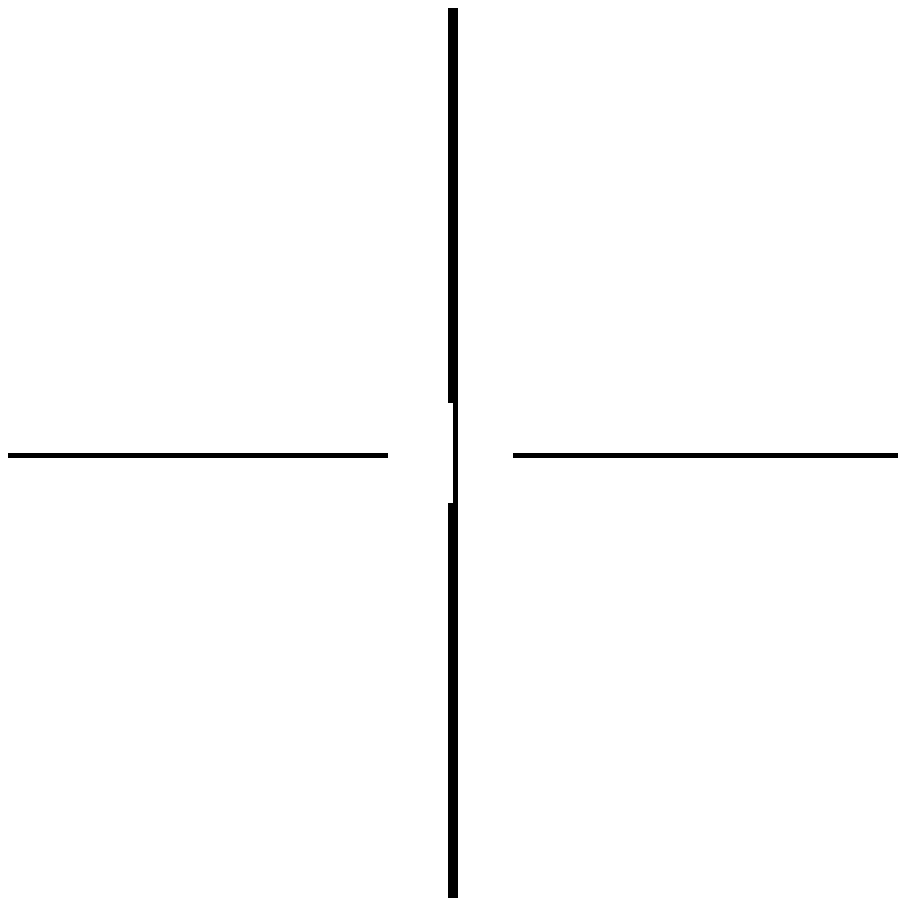}}
	  \Caption[artifacts][Artifacts in the amplitude spectra][{\figitem{a} The
	  log-amplitude-spectrum of the \emph{minimum 4-term Blackman-Harris} (B.H.)
	  window reveals a characteristic ``fingerprint'' (shown in this image), which
	  also emerges when averaging a big number of amplitude spectra of B.H.-windowed
	  faces. \figitem{b} The ``fingerprint'' is transformed into a binary image
	  by thresholding with $-0.25$ (black color indicates values with $1$,
	  and white indicates $0$). \figitem{c} Manually marked line artifacts which appear
	  by averaging the amplitude spectra of a big number of face images (here without windowing).}]
\end{figure*}
\begin{figure*}[htbp]
		  \emph{a}~\scalebox{0.3}{\includegraphics{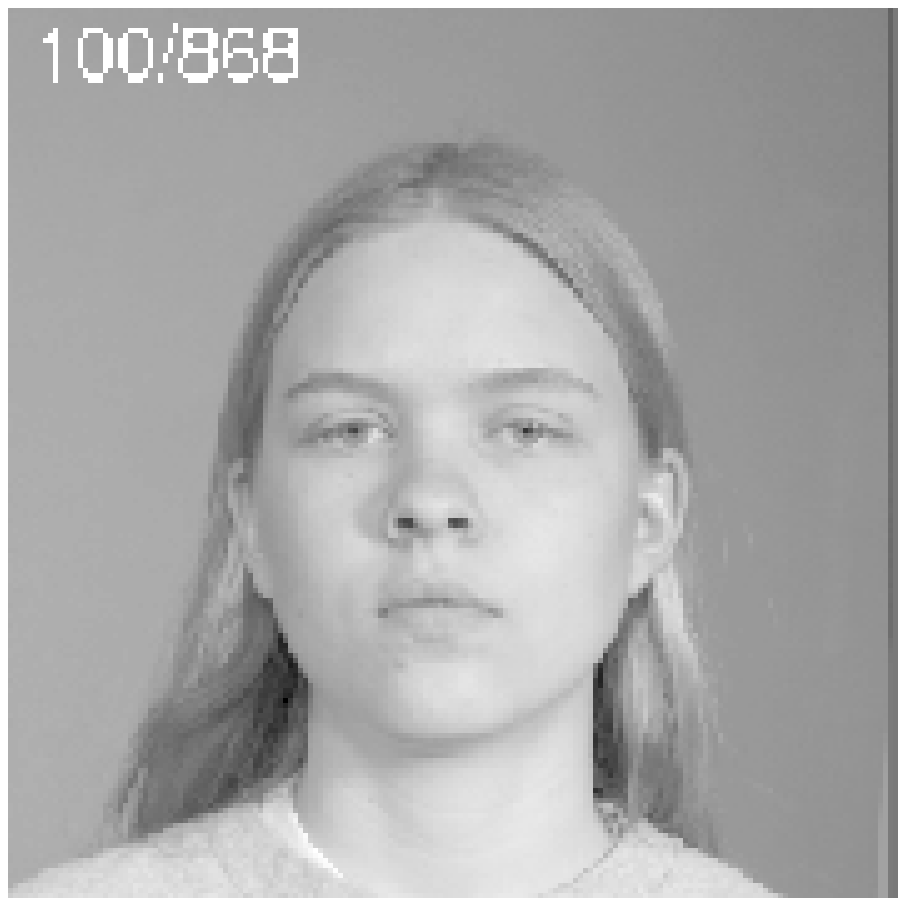}}
		  \emph{b}~\scalebox{0.3}{\includegraphics{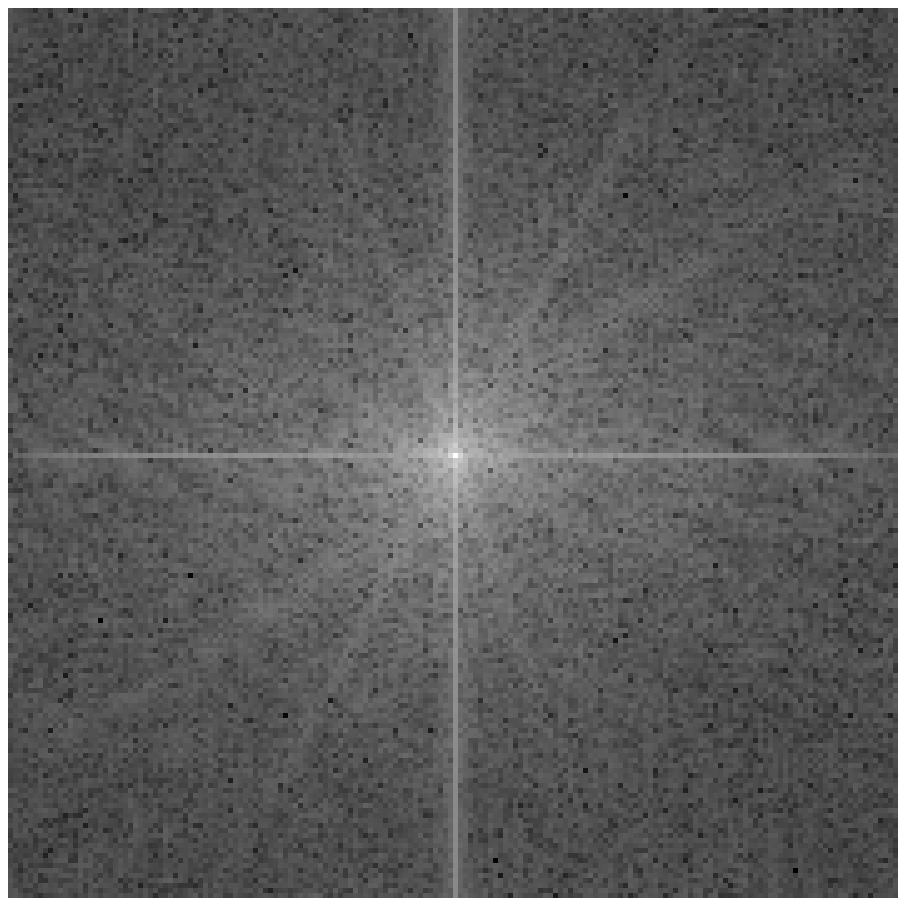}}
		  \emph{c}~\scalebox{0.3}{\includegraphics{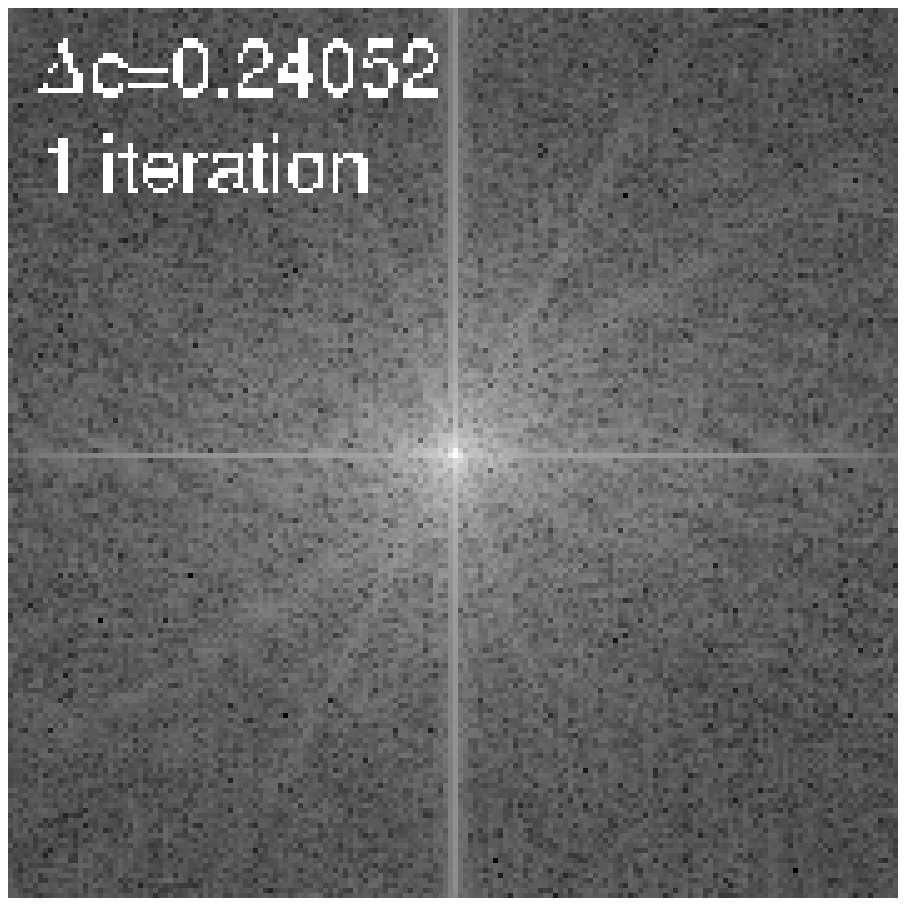}}
		  \emph{d}~\scalebox{0.3}{\includegraphics{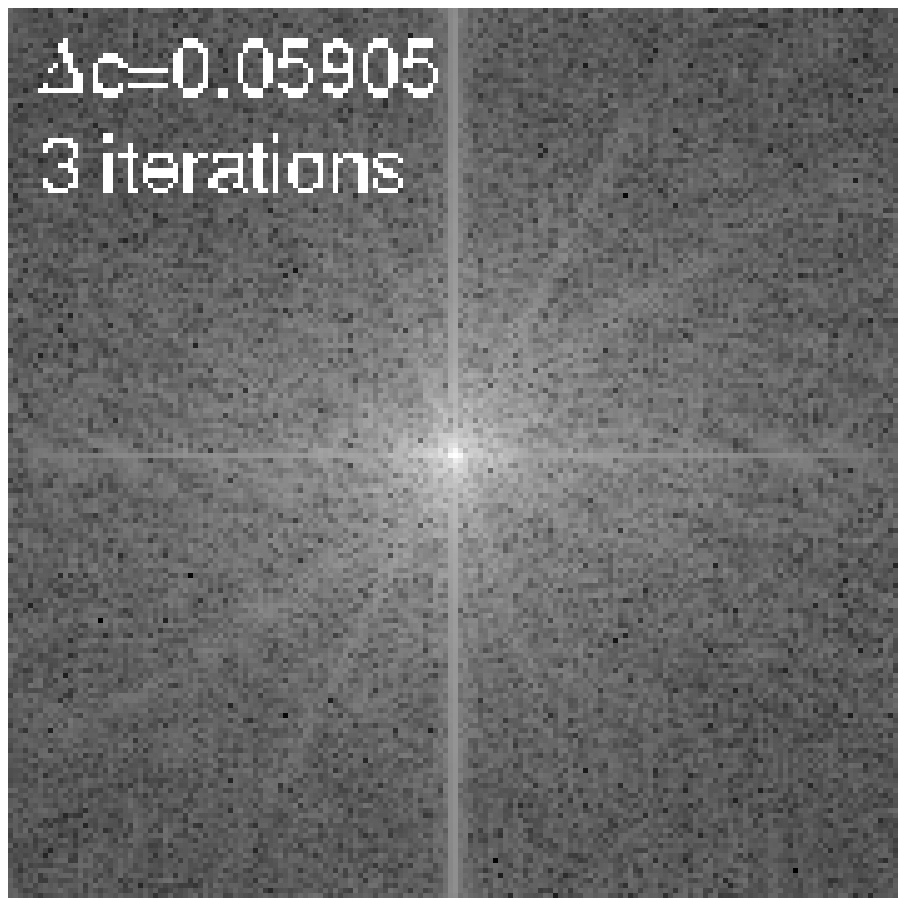}}
		  \emph{e}~\scalebox{0.3}{\includegraphics{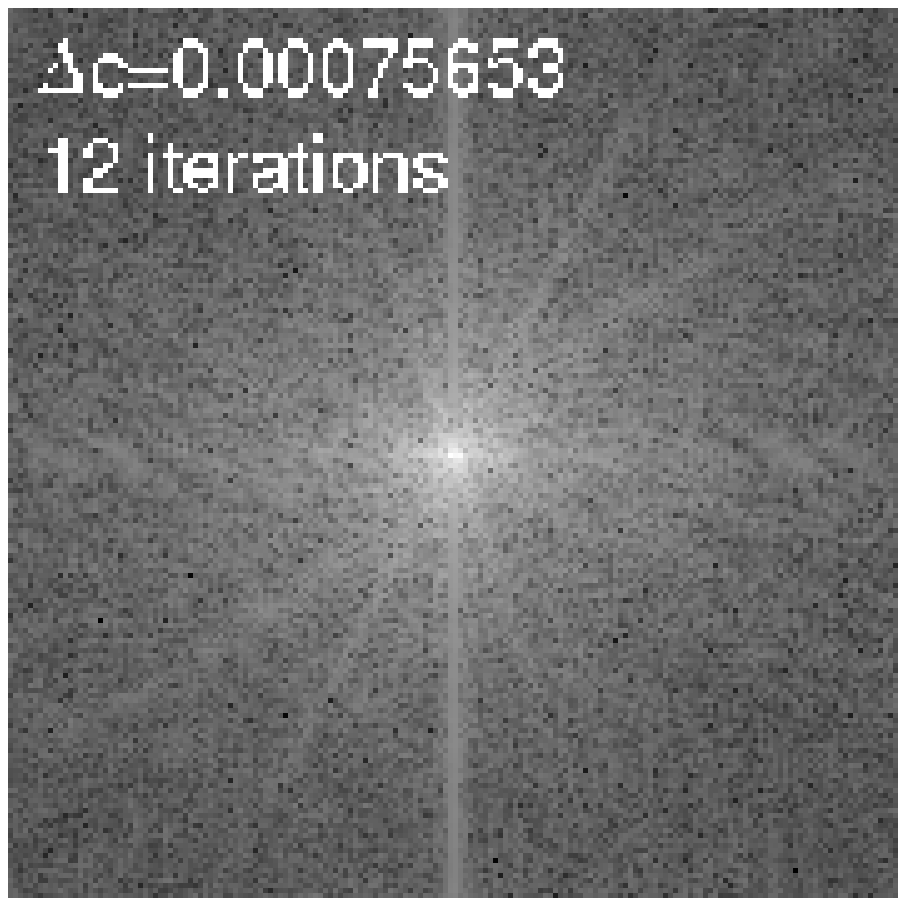}}
	  \Caption[InwardDiffusionDemo][Suppressing artifacts in the amplitude spectra][{This figure
	  illustrates how artifacts in the amplitude spectra are suppressed by a nonlinear
	  diffusion process, where the thresholded images of \suppfigure{artifacts} served
	  as spatially variant diffusion coefficient (see methods section).
	  \figitem{a} Original face image.
	  \figitem{b} The log-amplitude-spectrum of the image has horizontal and vertical
	  lines which are generated as a consequence of truncating the shoulder region
	  (c.f. \suppfig{artifacts}\figlabel{c}).
	  \figitem{c} The spectrum after one iteration of nonlinear diffusion, with a
	  difference in correlation to the original spectrum $\Delta c(1) \equiv c(0)-c(1)=0.24052$. 
	  The spurious lines are already attenuated.
	  \figitem{d} Three iterations with $\Delta c(3)=0.05905$.
	  \figitem{e} $12$ iterations with $\Delta c(12)<0.001$, which is the stopping criterion. 
	  The artificial lines are largely suppressed.  The rest of the amplitude spectrum
	  remains intact, and more interesting structures are now visible.}]
\end{figure*}
\begin{figure*}[htbp]
		  \emph{a}~\scalebox{0.375}{\includegraphics{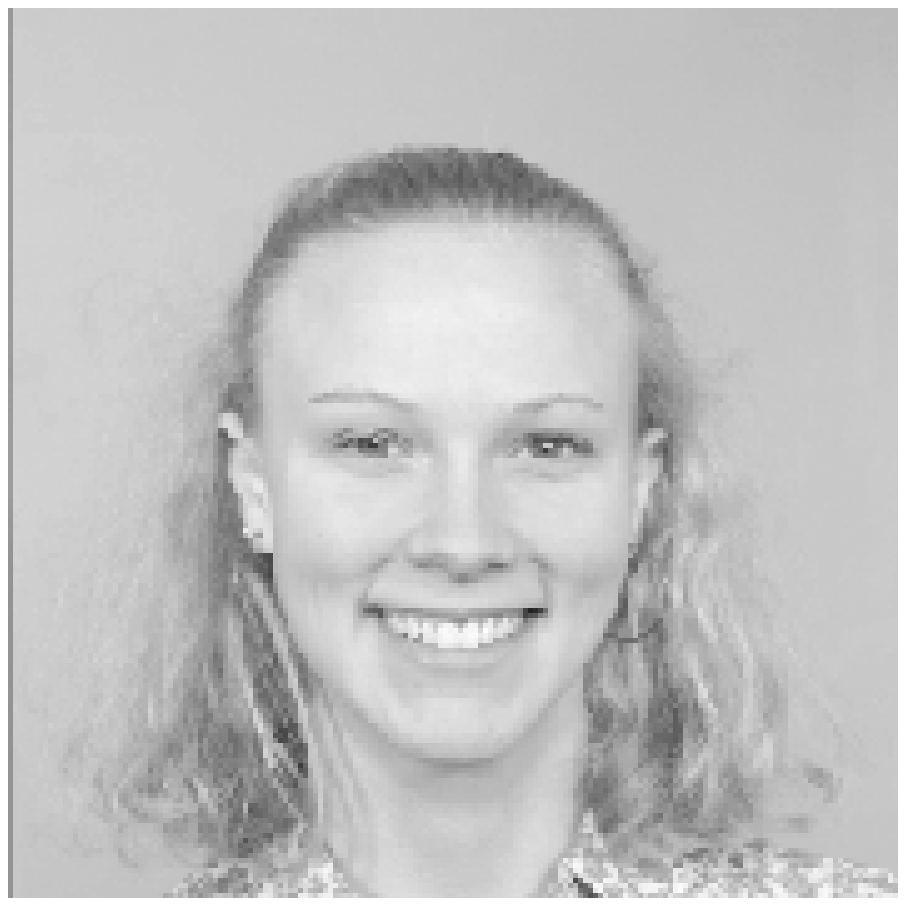}}
		  \emph{b}~\scalebox{0.375}{\includegraphics{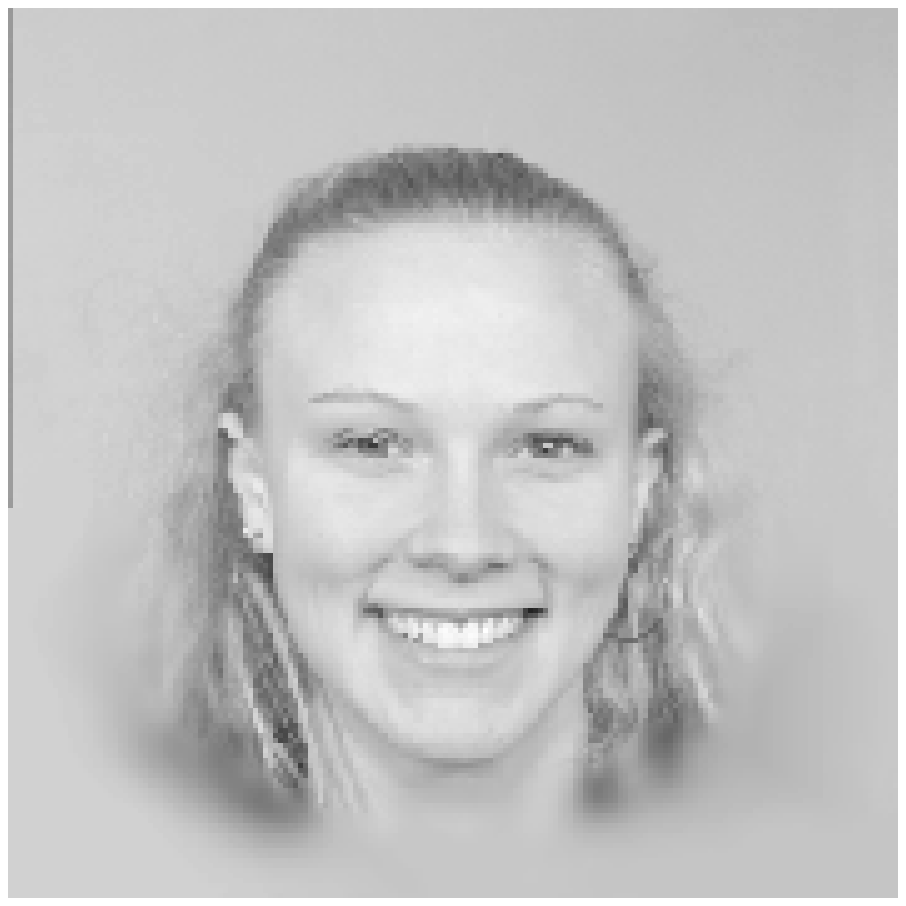}}
		  \emph{c}~\scalebox{0.375}{\includegraphics{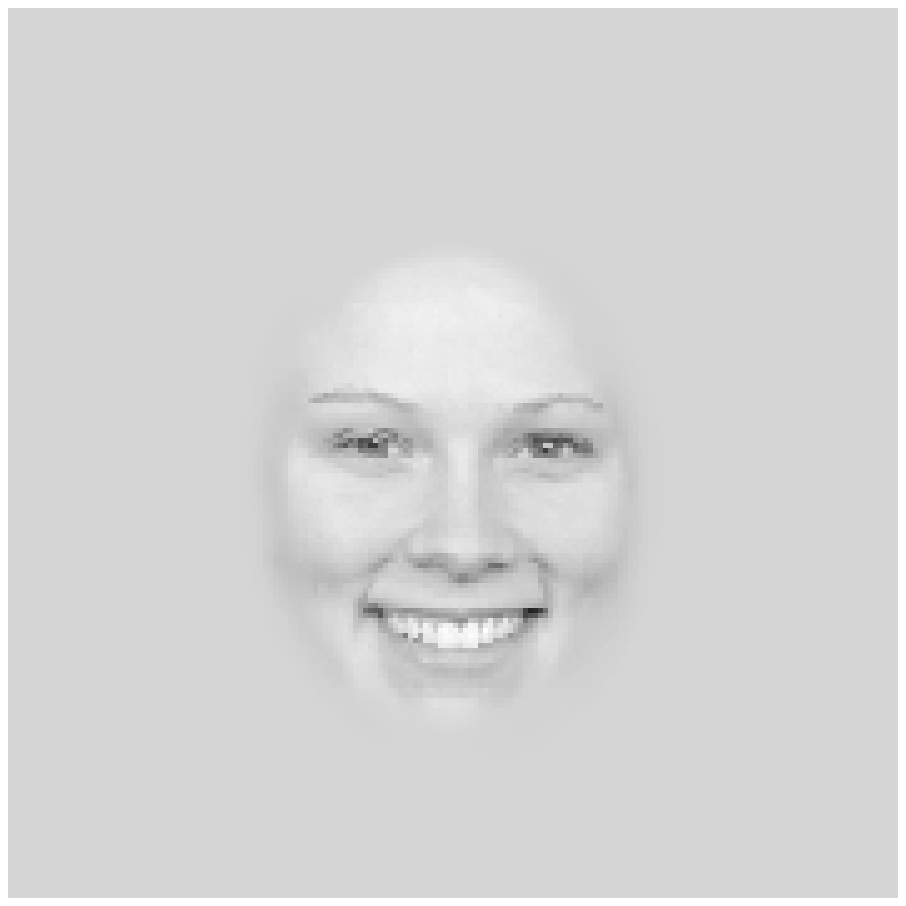}}
		  \emph{d}~\scalebox{0.375}{\includegraphics{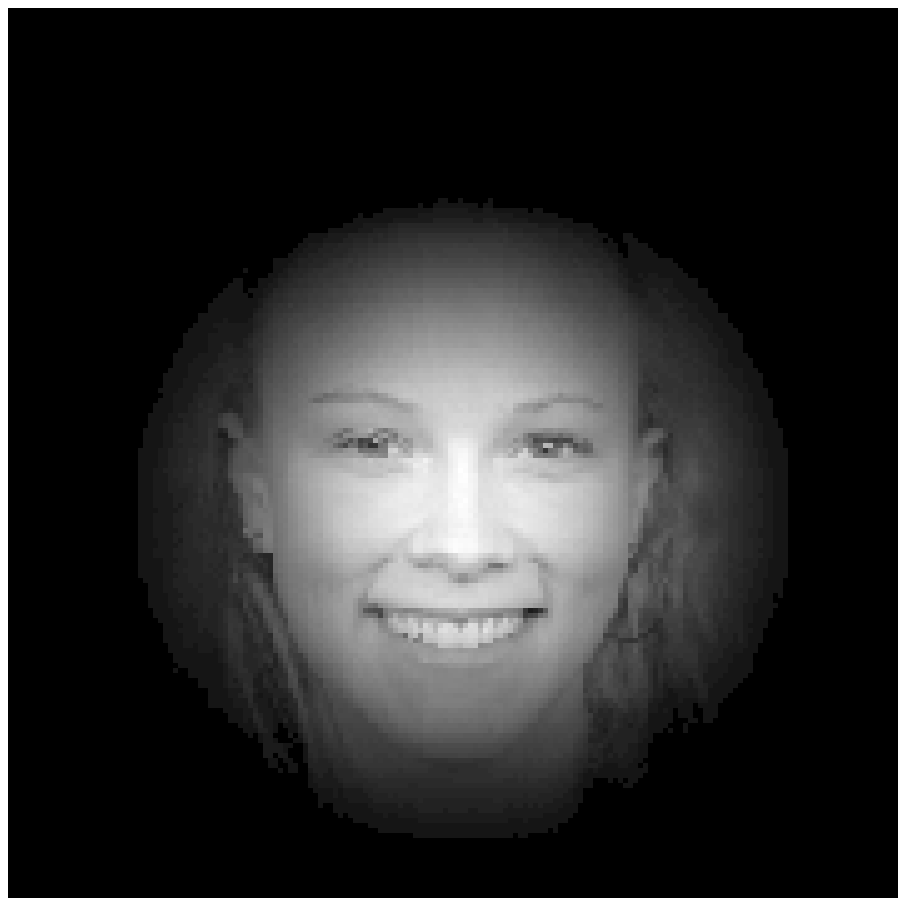}}\\

		  \emph{e}~\scalebox{0.375}{\includegraphics{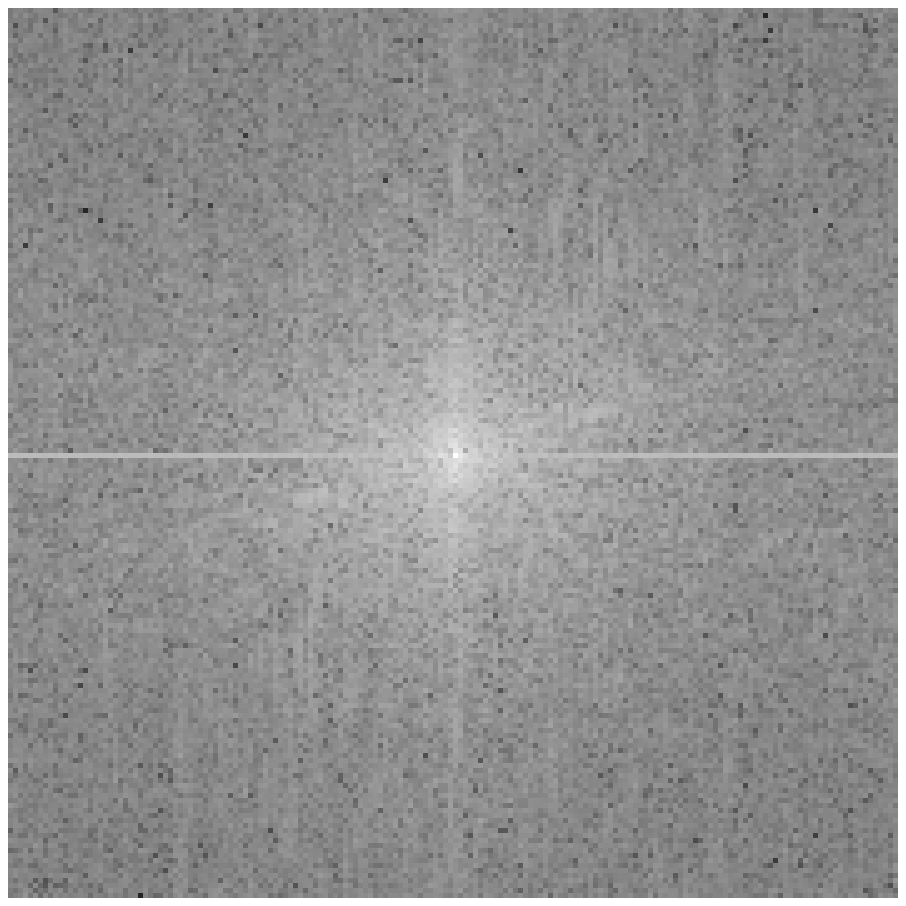}}
		  \emph{f}~\scalebox{0.375}{\includegraphics{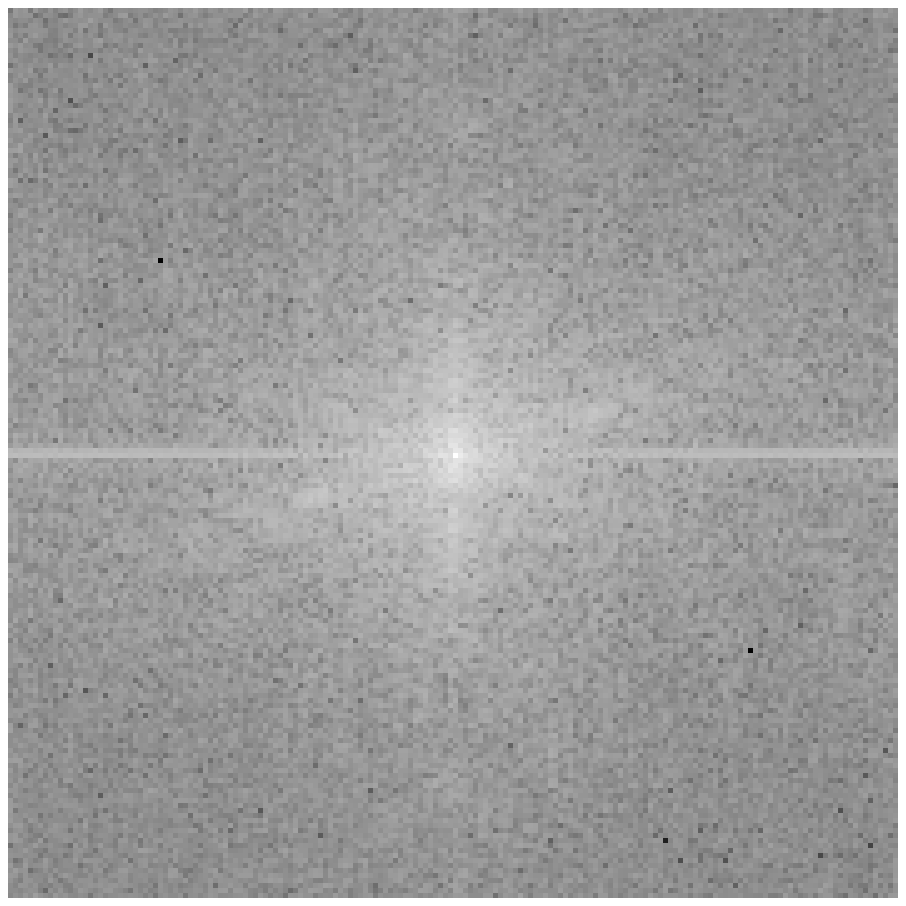}}
		  \emph{g}~\scalebox{0.375}{\includegraphics{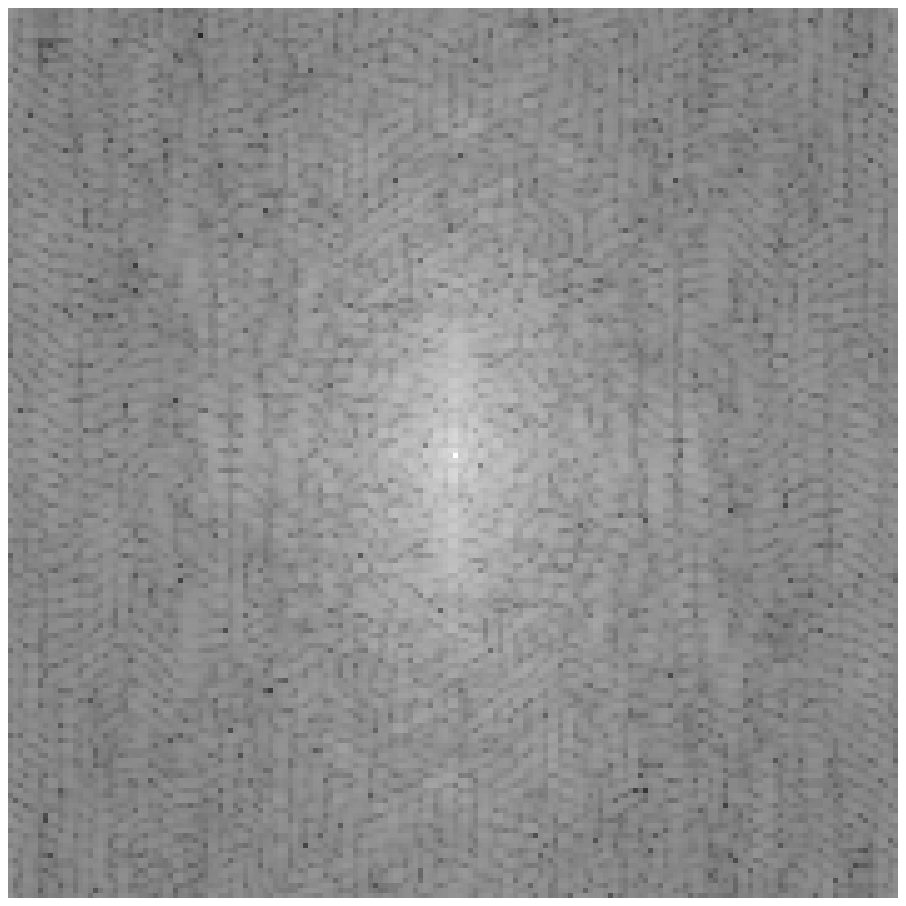}}
		  \emph{h}~\scalebox{0.375}{\includegraphics{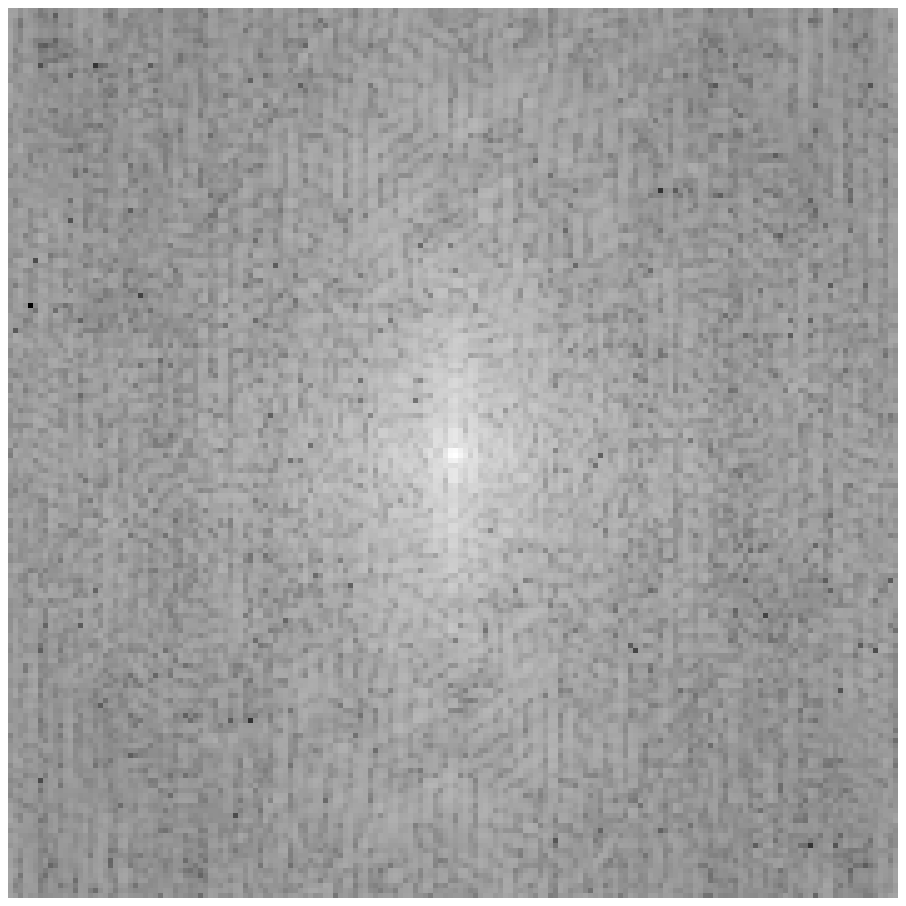}}

	  \Caption[MoonFaces][Suppression of external face features][{The images in the bottom row
	  \figitem{e--h} show the logarithmized amplitude spectra of the images \figitem{a--d}
	  (face ID 104).  The amplitude spectrum \figitem{e} of the original image \figitem{a}
	  	  shows spurious horizontal and the vertical lines.\figitem{b} The spurious vertical
	  line disappeared in the amplitude spectrum \figitem{f} when the shoulder region was
	  manually erased, and the horizontal line then had a smaller amplitude.
	  \figitem{c} Erasing all external face features led to the creation of a ``moonface'',
	  thereby suppressing all of the artificial lines \figitem{g}.
	  Finally, in \figitem{d}, a \emph{minimum 4-term Blackman-Harris window} was
	  centered at the nose position of the original image.  The corresponding
	  amplitude spectrum \figitem{h} of the windowed image is very similar to the amplitude
	  spectrum of the ``moonface'' spectrum (but see \suppfig{fems_testMI}).}]
\end{figure*}
\begin{figure*}[htbp]
	  \figitem{a}\scalebox{0.425}{\includegraphics{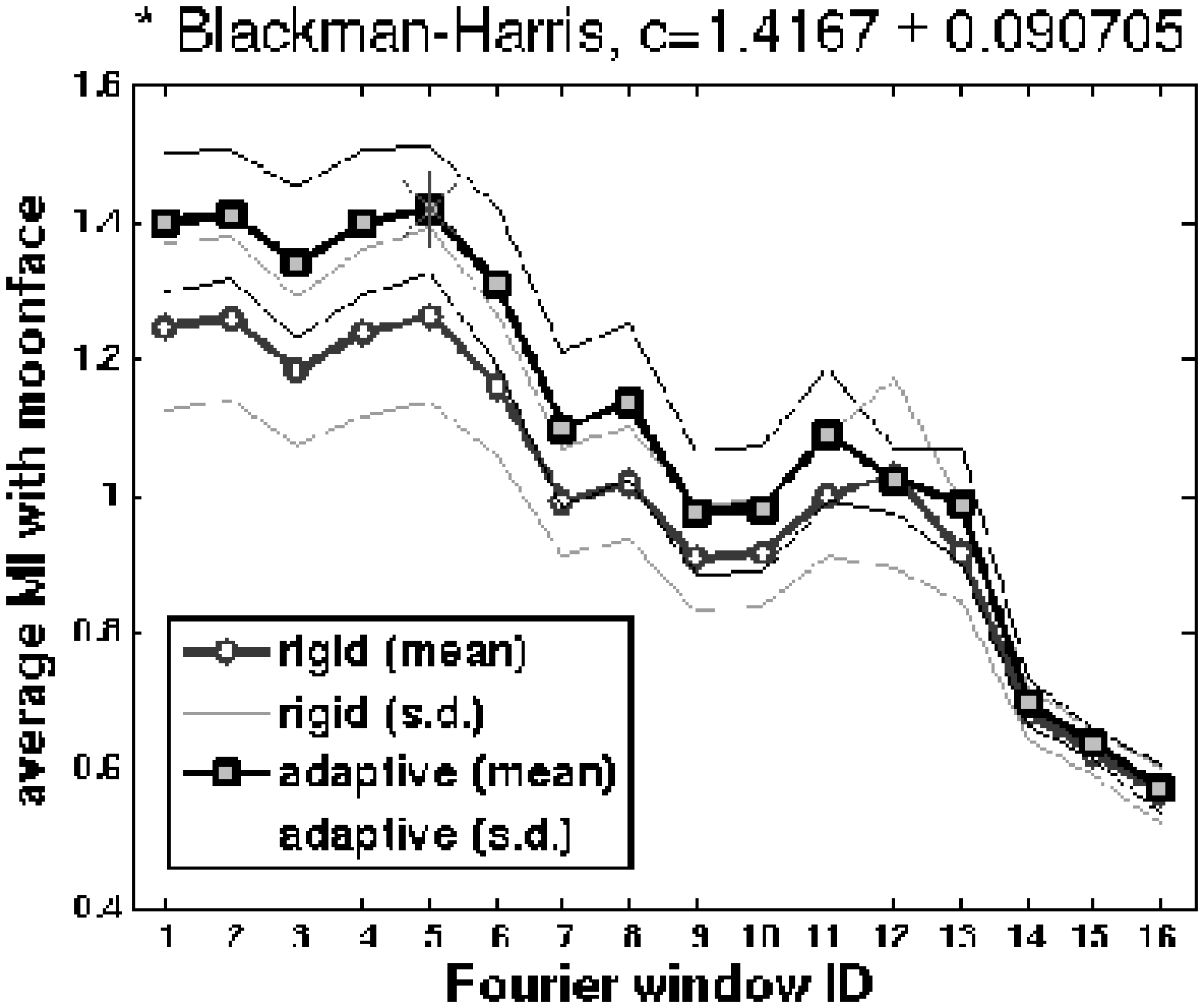}}
	  \Caption[fems_testMI][Similarities between the logarithmized amplitude spectra of
	  ``moonfaces'' and windowed faces][{For six selected female images
	  which revealed strong line artifacts in their amplitude
	  spectra, I computed similarity measures between the logarithmized amplitude
	  spectra of corresponding ``moonfaces'' (e.g., \suppfig{MoonFaces}\figlabel{g})
	  and windowed faces (e.g., \suppfig{MoonFaces}\figlabel{h}; the window type is
	  specified by the numbers at the abscissae). The plot shows mutual information
	  averaged across the the six images (mean
	  $\pm$ s.d.).  The center of the window was either positioned always at the
	  center position of each image (``rigid''), or at the nose position with
	  variable radius (``adaptive'') -- see legend.
	  The \emph{minimum 4-term Blackman-Harris window} scored the highest
	  similarity (indicated by a red star).
	  With correlation instead of mutual information, the curves show nearly the same
	  relative similarities.  In that case, the maximum average correlation value
	  ($\pm$ s.d.) was $0.87\pm0.02$ again for the adaptive \emph{minimum 4-term
	  Blackman-Harris window}.\newline
	  The identification numbers (``Fourier-IDs'') of the windows were
	  1=\emph{Chebyshev window}, 2=\emph{Nuttall-defined minimum 4-term Blackman-Harris window},
	  3=\emph{Bohman window}, 4=\emph{Parzen (de la Valle-Poussin) window}, 5=\emph{minimum 4-term
	  Blackman-Harris window}, 6=\emph{Blackman window}, 7=\emph{modified Bartlett-Hann window},
	  8=\emph{Hann (Hanning) window,} 9=\emph{triangular window}, 10=\emph{Bartlett window},
	  11=\emph{Gaussian window}, 12=\emph{flat top weighted window}, 13=\emph{Hamming window},
	  14=\emph{Tukey (tapered cosine) window}, 15=\emph{Kaiser window}, 16=\emph{sharp-edged disk}.}]
\end{figure*}
\begin{figure*}[htbp]
		\figitem{a}\scalebox{0.4}{\includegraphics{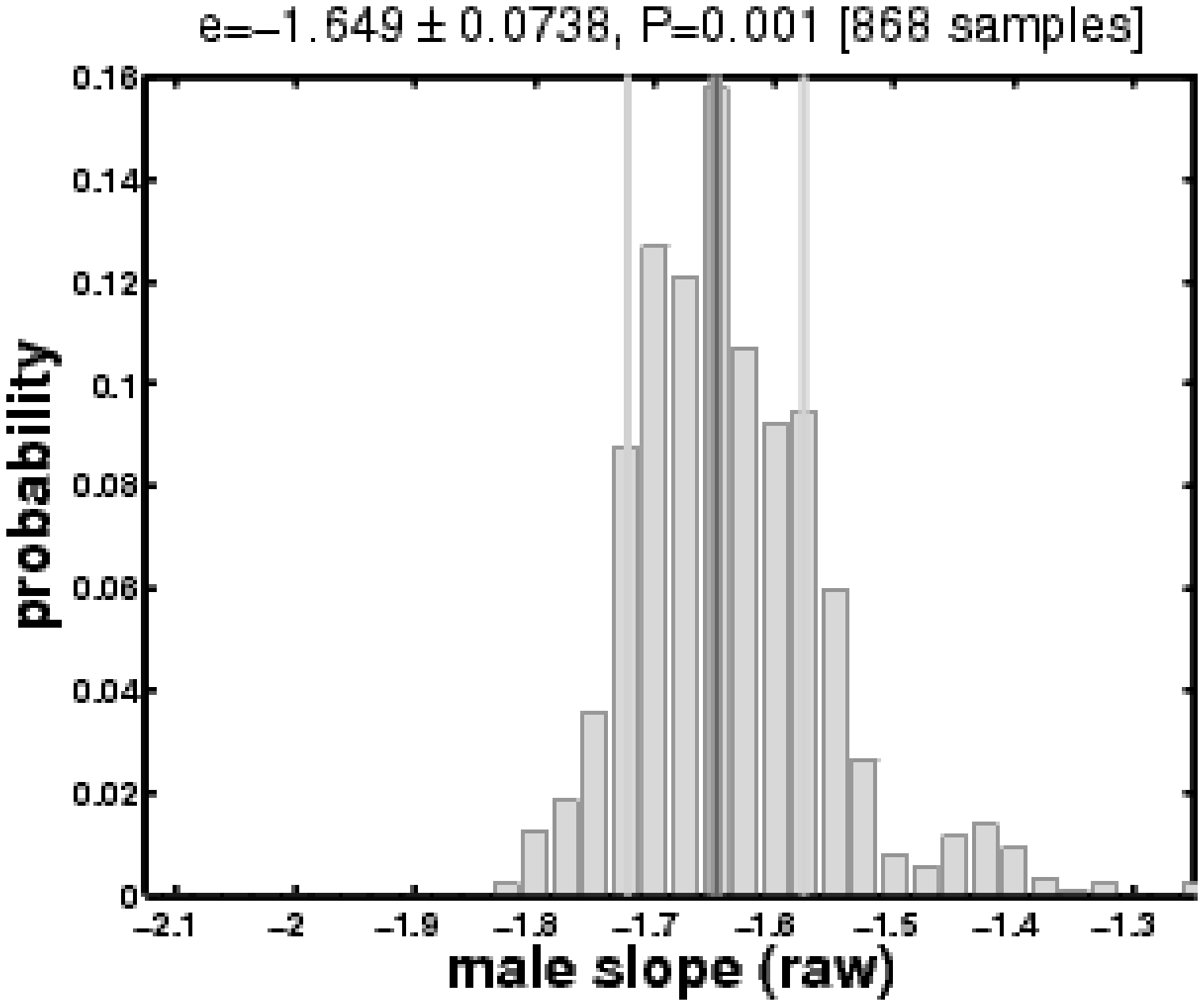}}
		\figitem{b}\scalebox{0.4}{\includegraphics{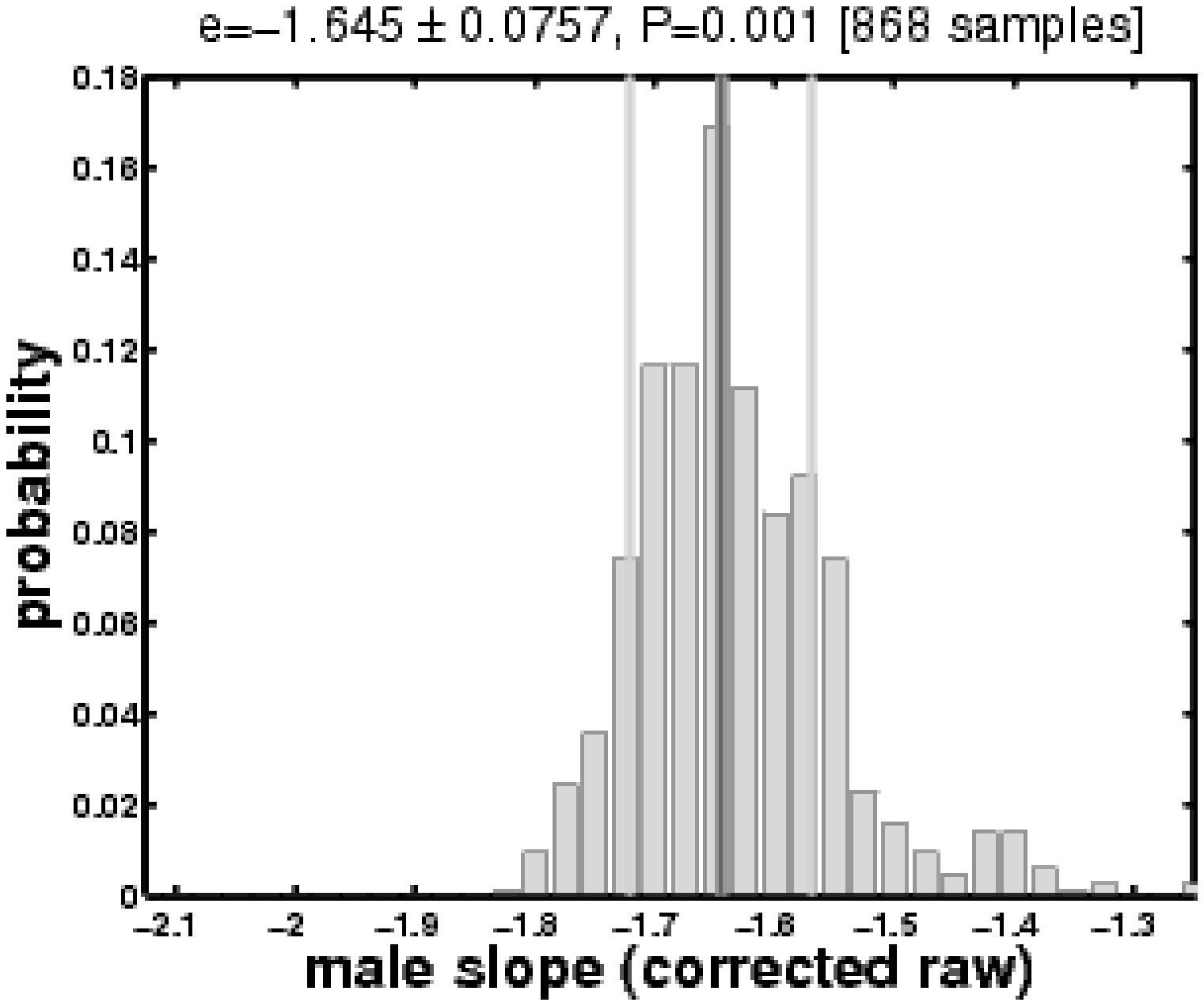}}\\
		\figitem{c}\scalebox{0.4}{\includegraphics{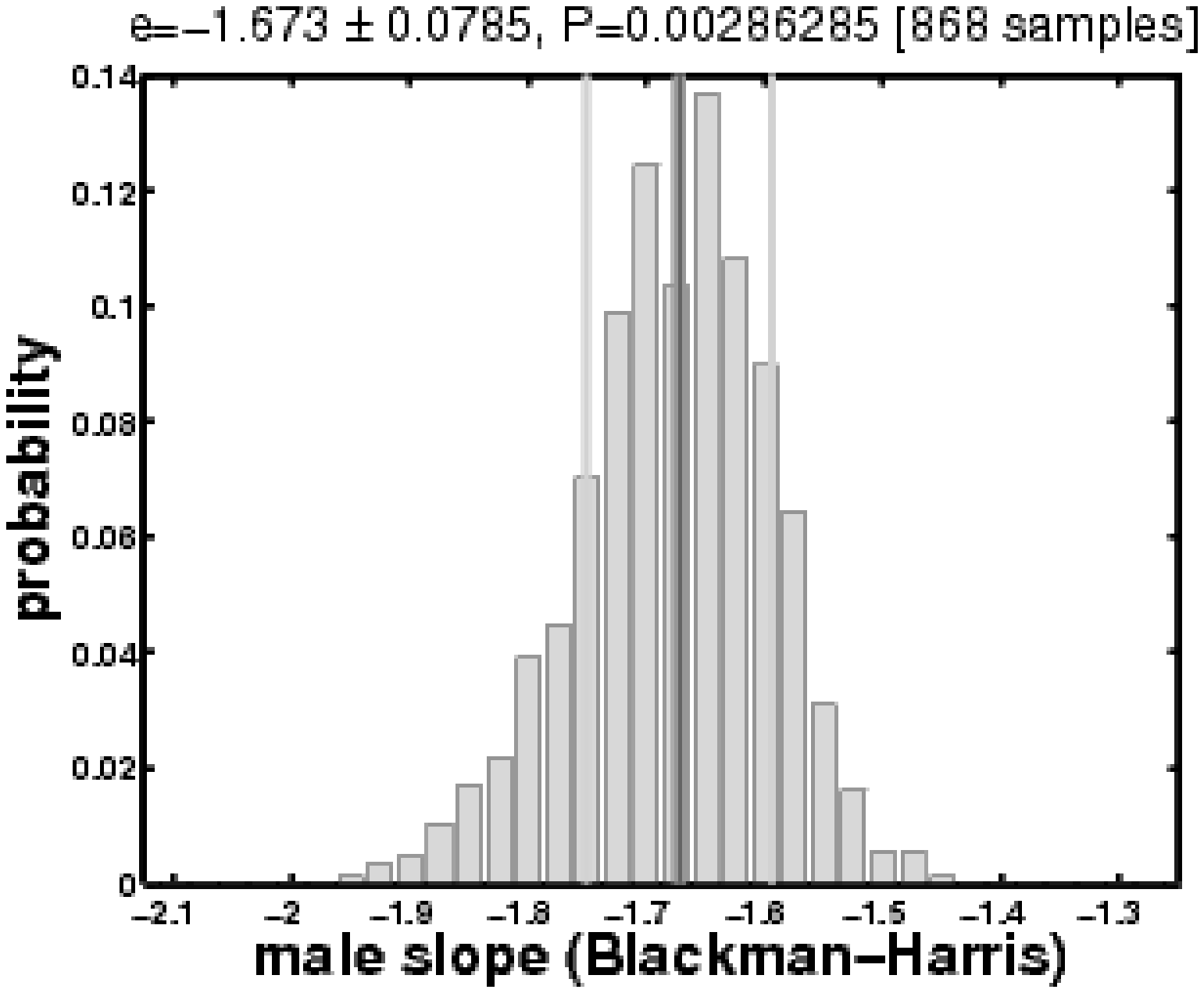}}
		\figitem{d}\scalebox{0.4}{\includegraphics{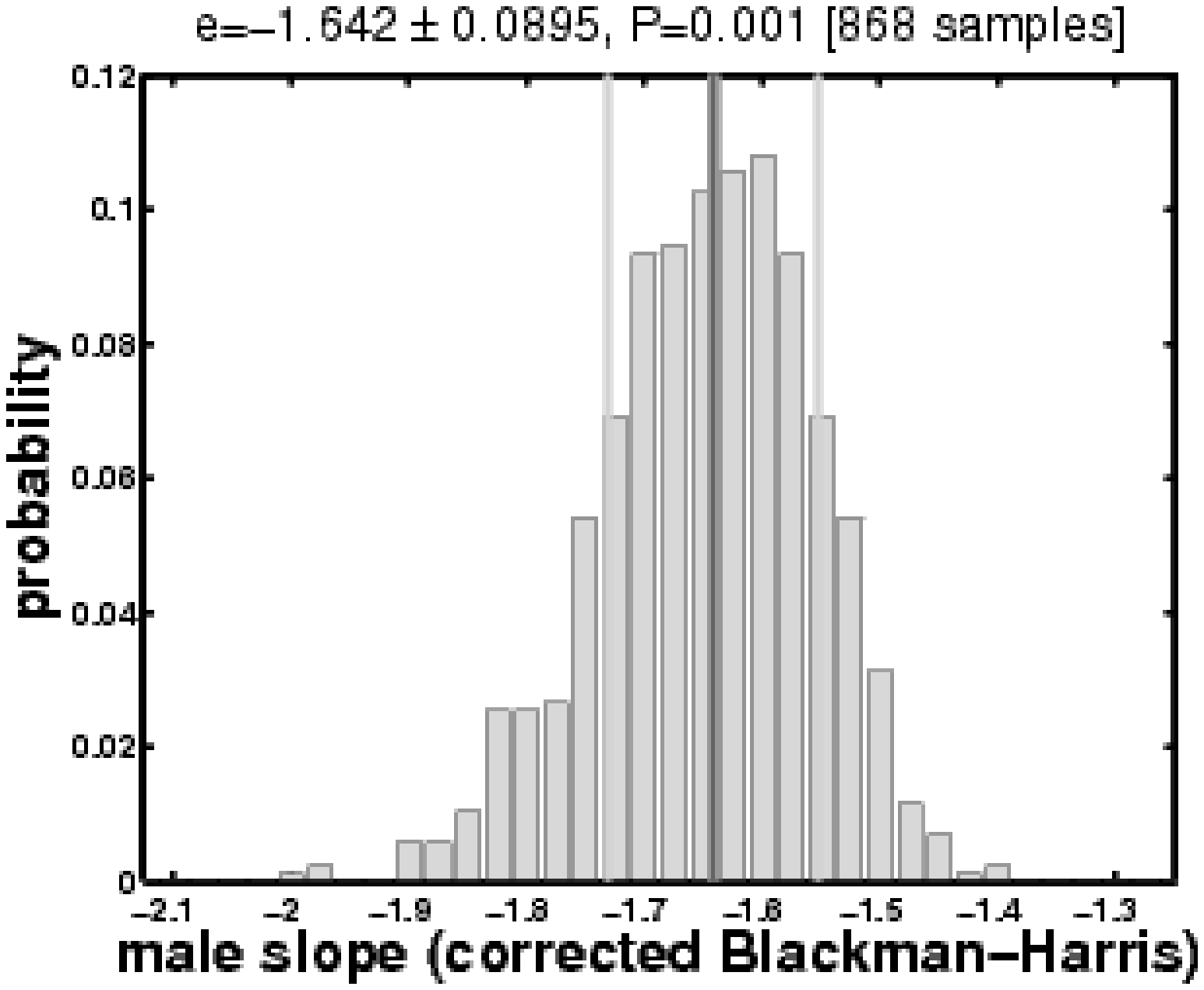}}
	\Caption[isoslopes_males][Slopes for individual images (male faces)][{The histograms
	show the probability of occurrence of slope values $e\equiv\alpha$ across all
	$868$ male face images.  For each face image, a corresponding slope value was obtained
	from fitting a line to the double-logarithmic representation of its isotropic 1-D amplitude
	spectrum (frequency range for fitting from $8$ to $100$ cycles per image).
	The centered vertical line in each histogram is the average $\alpha$, and the flanking
	lines denote $\pm 1$ s.d., respectively.  A Jarque-Bera test was used to test the
	slope values for normal distribution (this test could be applied because of our large
	sample size) -- corresponding $P$-values are indicated with each histogram.
	Corresponding histograms for female images are similar.
	\figitem{a} Raw spectrum: $\alpha=-1.649\pm0.0738$, $P<0.001$.
	\figitem{b} Corrected raw: $\alpha=-1.645\pm0.0757$, $P<0.001$.
	\figitem{c} Blackman-Harris: $\alpha=-1.673\pm0.0785$, $P=0.03$.
	\figitem{d} Corrected Blackman-Harris:  $\alpha=-1.642\pm0.0895$, $P<0.001$.}]
\end{figure*}
\begin{figure*}[bpth!]
	  \figitem{a}\scalebox{0.425}{\includegraphics{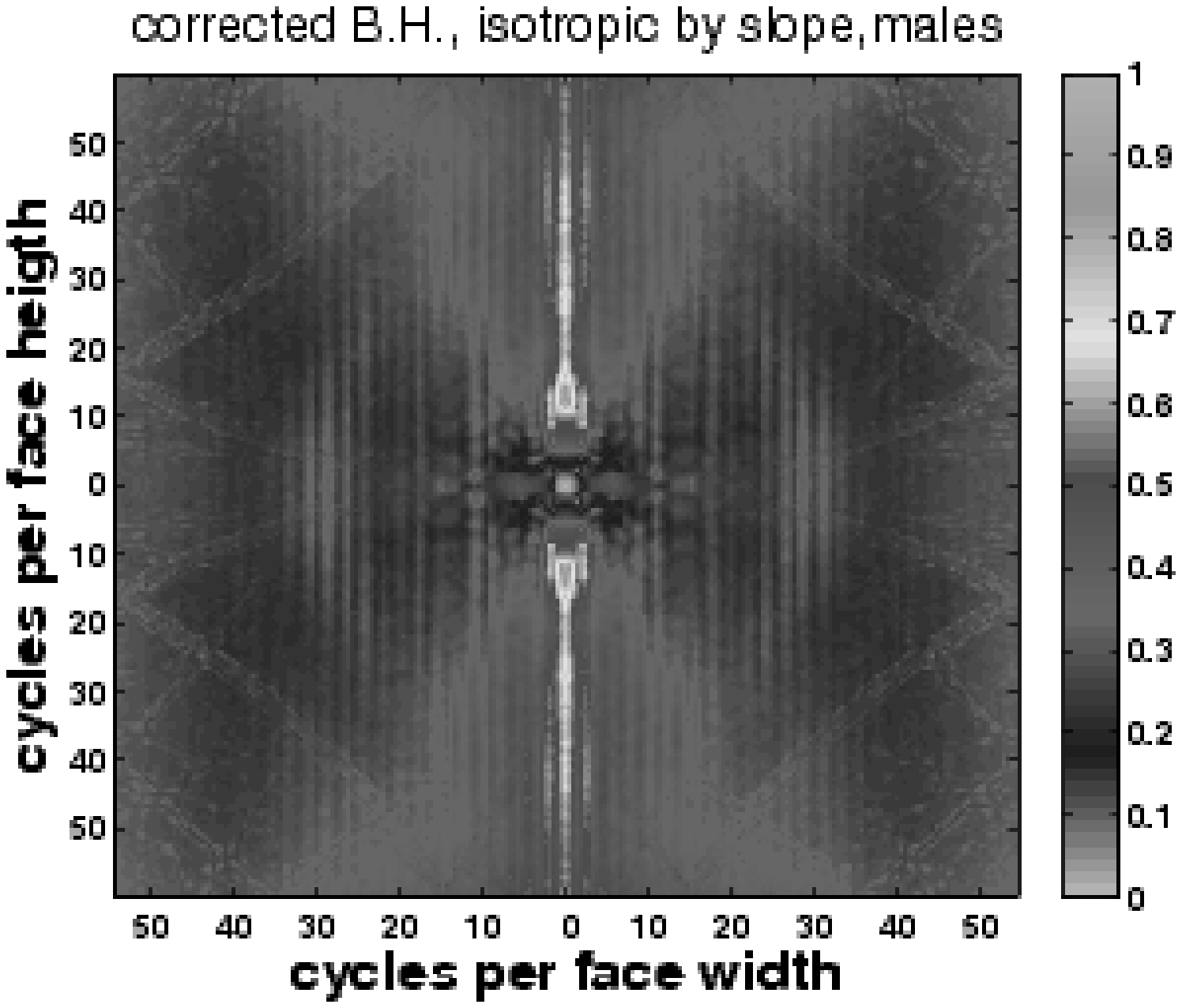}}
	  \figitem{b}\scalebox{0.425}{\includegraphics{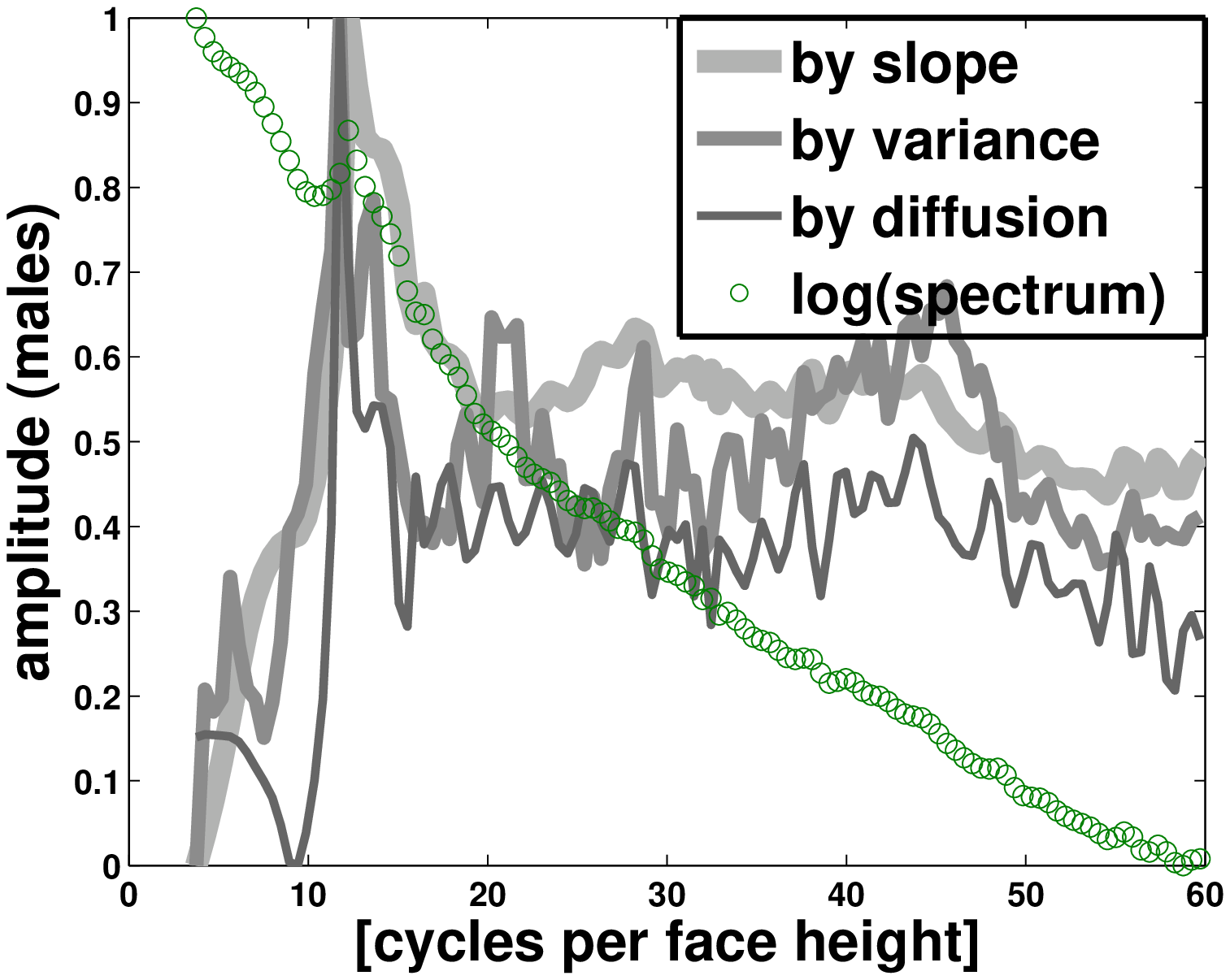}}
	  \Caption[WhiteMales2D][Whitening by slope][{Analogous to \fig{WhiteFemales2D},
	  but for face images of males.}]
\end{figure*}
\begin{figure*}[bpth!]
	  \figitem{a}\scalebox{0.425}{\includegraphics{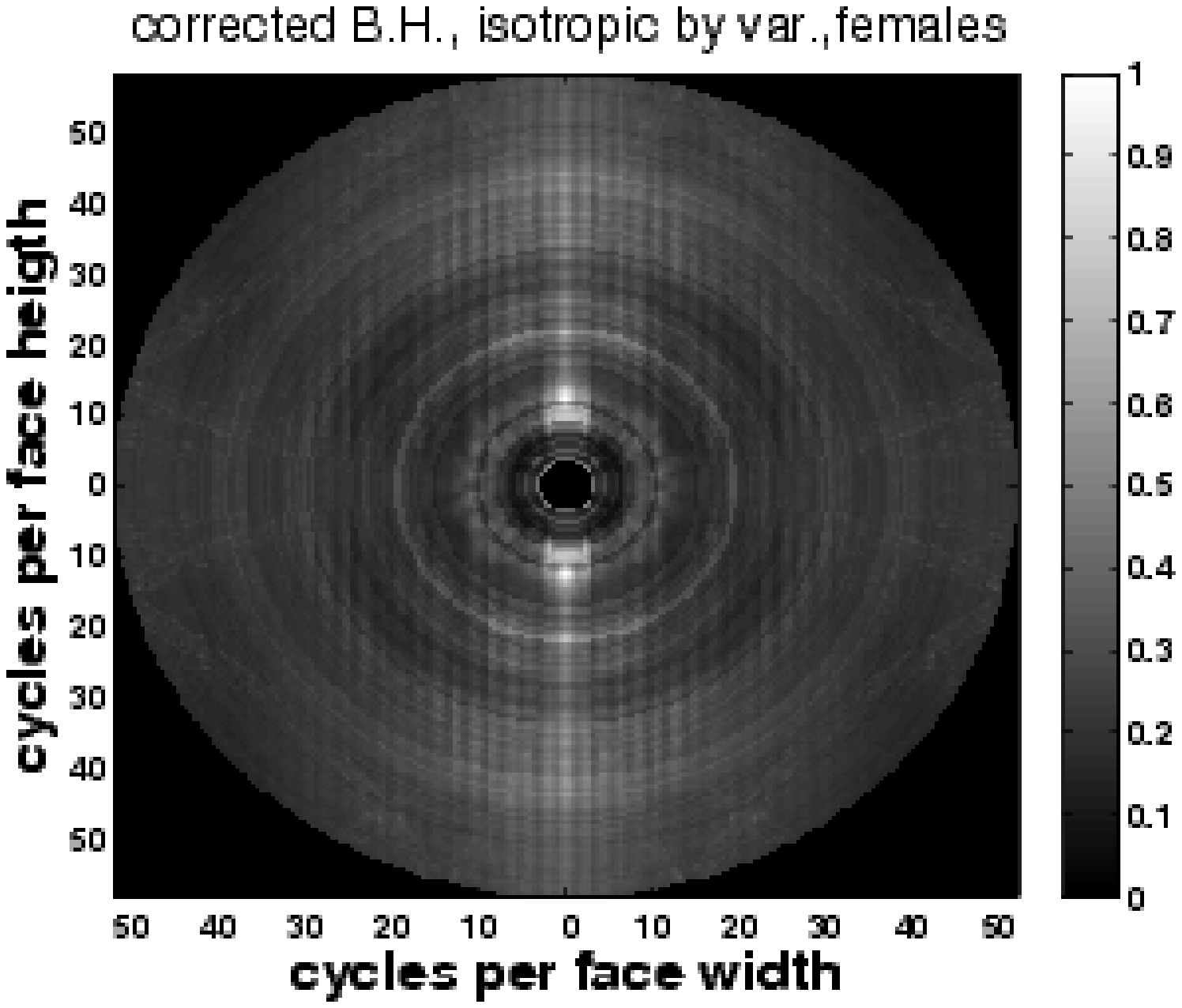}}
	  \figitem{b}\scalebox{0.425}{\includegraphics{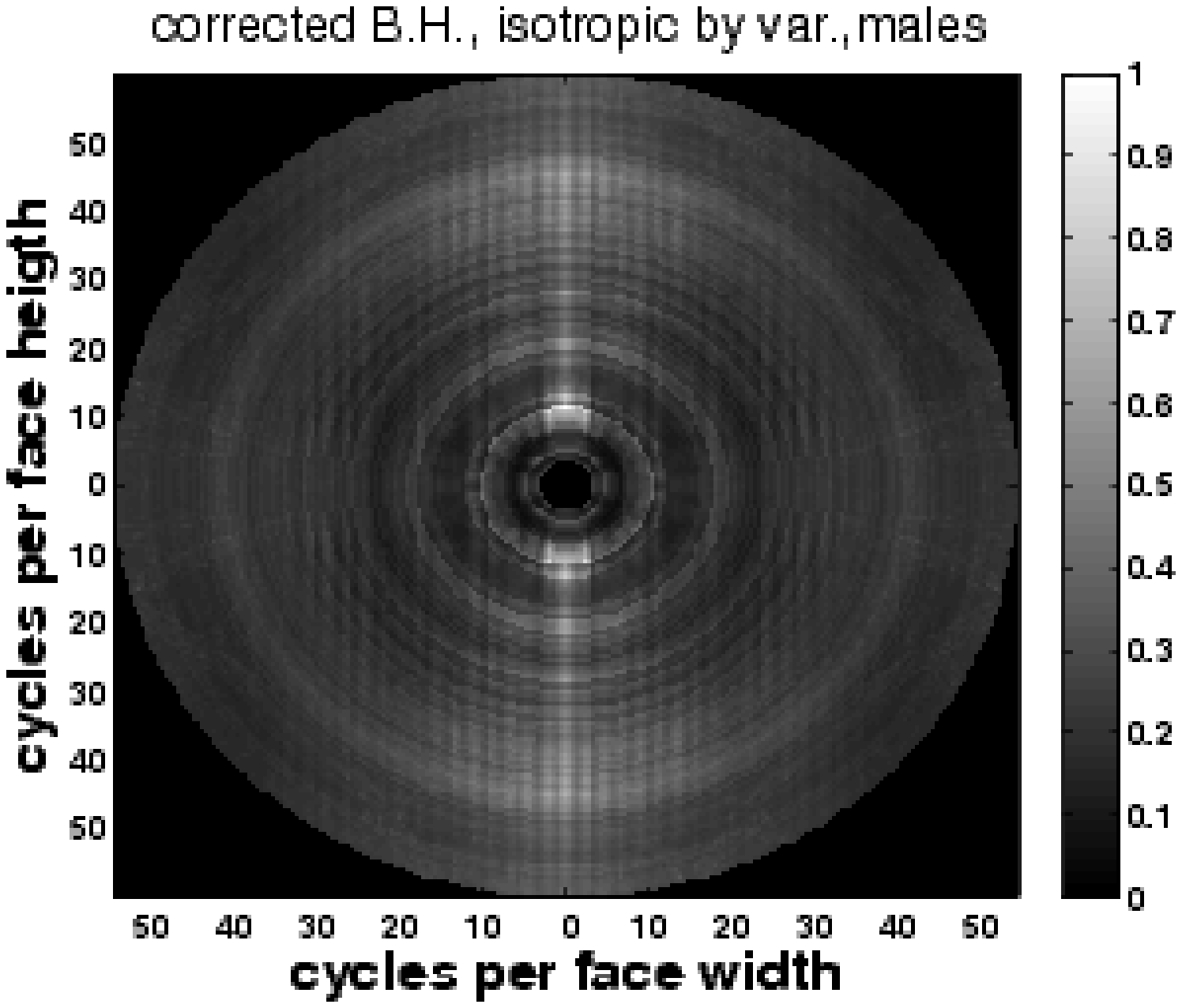}}
	  \Caption[VarWhitening][Whitening by variance][{Analogous to
	  \fig{WhiteFemales2D}\figitem{a} (females - left panel) and \fig{WhiteMales2D}\figitem{a}
	  (males - right panel) but here for variance-whitening.  Again, as with the slope-whitenend
	  spectra, maxima are revealed at low spatial frequencies for horizontally
	  oriented features (as indicated by the white regions close the center).}]
\end{figure*}
%
%

\begin{thebibliography}{}

\bibitem[Atick and Redlich, 1992]{Atick92a}
Atick, J. and Redlich, A. 1992.
\newblock What does the retina know about natural scenes?
\newblock {\em Neural Computation}, 4:196--210.

\bibitem[Attneave, 1954]{Attneave1954}
Attneave, F. 1954.
\newblock Some informational aspects of visual perception.
\newblock {\em Psychological Review}, 61(3):183--193.

\bibitem[Baddeley et~al., 1998]{BaddeleyEtAl98}
Baddeley, R., Abbott, L., Booth, M., Sengpiel, F., and Freeman, T. 1998.
\newblock Responses of neurons in primary and inferior temporal visual cortices
  to natural scenes.
\newblock {\em Proceedings of the Royal Society, London B}, 264:1775--1783.

\bibitem[Balboa and Grzywacz, 2001]{BalboaEtAl01}
Balboa, R. and Grzywacz, N. 2001.
\newblock Occlusions contribute to scaling in natural images.
\newblock {\em Vision Resarch}, 41:955--964.

\bibitem[Barlow, 1961]{Barlow61}
Barlow, H. 1961.
\newblock Possible principles underlying the transformation of sensory
  messages.
\newblock In Rosenblith, W., editor, {\em Sensory Communication}, pages
  217--234. MIT Press, Cambridge, MA.

\bibitem[Barlow, 1989]{Barlow89}
Barlow, H. 1989.
\newblock Unsupervised learning.
\newblock {\em Neural Computation}, 1:295--311.

\bibitem[Burton and Moorhead, 1987]{BurtonMoorehead87}
Burton, G. and Moorhead, I. 1987.
\newblock Color and spatial structure in natural scenes.
\newblock {\em Applied Optics}, 26(1):157--170.

\bibitem[Collin et~al., 2006]{CollinWangOByrne2006}
Collin, C., Wang, K., and O'Byrne, B. 2006.
\newblock Effects of image background on spatial-frequency threshold for face
  recognition.
\newblock {\em Perception}, 35:1459--1472.

\bibitem[Costen et~al., 1994]{CostenParkerCraw1994}
Costen, N., Parker, D., and Craw, I. 1994.
\newblock Spatial content and spatial quantisation effects in face recognition.
\newblock {\em Perception}, 23:129--146.

\bibitem[Costen et~al., 1996]{CostenParkerCraw1996}
Costen, N., Parker, D., and Craw, I. 1996.
\newblock Effects of high-pass and low-pass spatial filtering on face
  identification.
\newblock {\em Perception and Psychophysics}, 58:602--612.

\bibitem[Field, 1987]{Field87}
Field, D. 1987.
\newblock Relations between the statistics of natural images and the response
  properties of cortical cells.
\newblock {\em Journal of the Optical Society of America A}, 4(12):2379--2394.

\bibitem[Fiorentini et~al., 1983]{FiorentiniMaffeiSandini1983}
Fiorentini, A., Maffei, L., and Sandini, G. 1983.
\newblock The role of high spatial frequencies in face perception.
\newblock {\em Perception}, 12:195--201.

\bibitem[Ginsburg, 1978]{Ginsburg1978}
Ginsburg, A. 1978.
\newblock {\em Visual information processing based on spatial filters
  constrained by biological data}.
\newblock PhD thesis, Cambridge University, Cambridge, England.

\bibitem[Graham et~al., 2006]{GrahamChandlerField2006}
Graham, D., Chandler, D., and Field, D. 2006.
\newblock Can the theory of ``whitening'' explain the center-surround
  properties of retinal ganglion cell receptive fields?
\newblock {\em Vision Research}, 46(18):2901--2913.

\bibitem[Harris, 1978]{Harris78}
Harris, F. 1978.
\newblock On the use of windows for harmonic analysis with the discrete
  {F}ourier transform.
\newblock {\em Proceedings of the IEEE}, 66(1):51--84.

\bibitem[Hayes et~al., 1986]{HayesMorroneBurr1986}
Hayes, A., Morrone, M., and Burr, D. 1986.
\newblock Recognition of positive and negative band-pass filtered images.
\newblock {\em Perception}, 15:595--602.

\bibitem[Hosoya et~al., 2005]{HosoyaEtAl05}
Hosoya, T., Baccus, S., and Meister, M. 2005.
\newblock Dynamic predictive coding by the retina.
\newblock {\em Nature}, 436:71--77.

\bibitem[Laughlin et~al., 1998]{LaughlinEtAl98}
Laughlin, S., de~Ruyter~van Steveninck, R., and Anderson, J. 1998.
\newblock The metabolic cost of neural information.
\newblock {\em Nature Neuroscience}, 1(1):36--41.

\bibitem[Laughlin and Sejnowski, 2003]{LaughlinSejnowski03}
Laughlin, S. and Sejnowski, T. 2003.
\newblock Communication in neural networks.
\newblock {\em Science}, 301:1870--1874.

\bibitem[Lenny, 2003]{Lenny03}
Lenny, P. 2003.
\newblock The cost of cortical computation.
\newblock {\em Current Biology}, 13:493--497.

\bibitem[Levy and Baxter, 1996]{LevyBaxter96}
Levy, W. and Baxter, R. 1996.
\newblock Energy-efficient neural codes.
\newblock {\em Neural Computation}, 8:531--543.

\bibitem[Linsker, 1988]{Linsker1988}
Linsker, R. 1988.
\newblock Self-organization in a perceptual network.
\newblock {\em IEEE Transactions on Computer}, 21(3):105--117.

\bibitem[Nadal et~al., 1998]{NadalBrunelParga1998}
Nadal, J.-P., Brunel, N., and Parga, N. 1998.
\newblock Nonlinear feedforward networks with stochastic outputs: Infomax
  implies redundancy reduction.
\newblock {\em Network: Computation in Neural Systems}, 9:1--11.

\bibitem[N{\"a}s{\"a}nen, 1999]{Nasanen1999}
N{\"a}s{\"a}nen, R. 1999.
\newblock Spatial frequency bandwidth used in the recognition of facial images.
\newblock {\em Vision Research}, 39:3824--3833.

\bibitem[Ojanp{\"a}{\"a} and N{\"a}s{\"a}nen, 2003]{OjanpaaNasanen2003}
Ojanp{\"a}{\"a}, H. and N{\"a}s{\"a}nen, R. 2003.
\newblock Utilisation of spatial frequency information in face search.
\newblock {\em Vision Research}, 43(24):2505--2515.

\bibitem[Peli et~al., 1994]{PeliEtAl1994}
Peli, E., Lee, E., Trempe, C., and Buzney, S. 1994.
\newblock Image enhancement for the visually impaired: the effects of
  enhancement on face recognition.
\newblock {\em Journal of the Optical Society of America A}, 11:1929--1939.

\bibitem[Ruderman, 1997]{Ruderman97}
Ruderman, D. 1997.
\newblock Origins of scaling in natural images.
\newblock {\em Vision Research}, 37(23):3385--3398.

\bibitem[Srinivasan et~al., 1982]{SrinivasanLaughlinDubs82}
Srinivasan, M., Laughlin, S., and Dubs, A. 1982.
\newblock Predictive coding: a fresh view of inhibiton in the retina.
\newblock {\em Proceedings of the Royal Society of London B}, 216:427--459.

\bibitem[Switkes et~al., 1978]{SwitkesEtAl78}
Switkes, E., Mayer, M., and Sloan, J. 1978.
\newblock Spatial frequency analysis of the visual environment: anisotropy and
  the carpentered environment hypothesis.
\newblock {\em Vision Research}, 18:1393--1399.

\bibitem[Tieger and Ganz, 1979]{TiegerGanz1979}
Tieger, T. and Ganz, L. 1979.
\newblock Recognition of faces in the presence of two-dimensional sinusoidal
  masks.
\newblock {\em Perception and Psychophysics}, 26:163--167.

\bibitem[Tolhurst et~al., 1992]{TolhurstEtAl92}
Tolhurst, D., Tadmor, Y., and Chao, T. 1992.
\newblock Amplitude spectra of natural images.
\newblock {\em Ophthalmic and Physiological Optics}, 12:229--232.

\bibitem[Torralba and Oliva, 2003]{TorralbaOliva03}
Torralba, A. and Oliva, A. 2003.
\newblock Statistics of natural image categories.
\newblock {\em Network: Computation in Neural Systems}, 14:391--412.

\bibitem[van~der Schaaf and van Hateren, 1996]{SchaafHateren96}
van~der Schaaf, A. and van Hateren, J. 1996.
\newblock Modelling the power spectra of natural images: statistics and
  information.
\newblock {\em Vision Research}, 36(17):2759--2770.

\bibitem[Wainwright, 1999]{Wainwright99}
Wainwright, M. 1999.
\newblock Visual adaptation as optimal information transmission.
\newblock {\em Vision Research}, 39:3960--3974.

\bibitem[Wiener, 1964]{Wiener1964}
Wiener, N. 1964.
\newblock {\em Extrapolation, Interpolation, and Smoothing of Stationary Time
  Series}.
\newblock The MIT Press, Cambridge, Massachusetts.

\end{thebibliography}
\end{document}